\documentclass{article}

\usepackage[authoryear]{natbib}

\usepackage{arxiv}

\usepackage{moreverb}
\usepackage{caption}
\usepackage{subcaption}
\usepackage{array}
\usepackage{hyperref}
\usepackage{longtable}
\usepackage{tabu}
\usepackage[utf8]{inputenc} % allow utf-8 input
\usepackage[T1]{fontenc}    % use 8-bit T1 fonts
\usepackage{hyperref}       % hyperlinks
\usepackage{url}            % simple URL typesetting
\usepackage{booktabs}       % professional-quality tables
\usepackage{amsfonts}       % blackboard math symbols
\usepackage{nicefrac}       % compact symbols for 1/2, etc.
\usepackage{microtype}      % microtypography
\usepackage{lipsum}		% Can be removed after putting your text content
\usepackage{graphicx}
\usepackage{natbib}
\usepackage{doi}

\title{Human-Centred Learning Analytics and AI in Education: a Systematic Literature Review}                      

\author{
        Riordan Alfredo\thanks{Centre for Learning Analytics Monash - Monash University, 20 Exhibition Walk, Clayton, 3800, VIC, Australia}\\
        \texttt{\href{https://orcid.org/0000-0001-5440-6143}{\hspace{1mm}riordan.alfredo@monash.edu}}
 \\
	\And
	 Vanessa Echeverria*\thanks{Escuela Superior Politécnica del Litoral, 30.5 Via Perimetral, Guayaquil, Ecuador}\\ 
        \texttt{\href{https://orcid.org/0000-0002-2022-9588}{\hspace{1mm}vanessa.echeverria@monash.edu}}
 \\
	\AND Yueqiao Jin*\\
        \texttt{\href{https://orcid.org/0009-0003-7309-4984}{\hspace{1mm}ariel.jin@monash.edu}}
 \\
	\And Lixiang Yan*\\
        \texttt{\href{https://orcid.org/0000-0003-3818-045X}{\hspace{1mm}lixiang.yan@monash.edu}}
 \\
	\And Zachari Swiecki*\\
        \texttt{\href{https://orcid.org/0000-0002-7414-5507}{\hspace{1mm}zach.swiecki@monash.edu}}
 \\
	\AND Dragan Gašević*\\
        \texttt{\href{https://orcid.org/0000-0001-9265-1908}{\hspace{1mm}dragan.gasevic@monash.edu}}
 \\
	\And Roberto Martinez-Maldonado*\\
        \texttt{\href{https://orcid.org/0000-0002-8375-1816}{\hspace{1mm}roberto.martinezmaldonado@monash.edu}}
 \\
}

\hypersetup{
pdftitle={Human-Centred Learning Analytics and AI in Education: a Systematic Literature Review},
pdfsubject={cs.HCI},
pdfauthor={Riordan Alfredo, Vanessa Echeverria, Yueqiao Jin, Lixiang Yan, Zachari Swiecki, Dragan Gasevic, Roberto Martinez-Maldonado},
pdfkeywords={Human-centered AI, Human-centered learning analytics, AI in education, Stakeholders involvement, Education technology, Ethical considerations},
}

\date{}

\begin{document}
\maketitle

% Here goes the abstract
\begin{abstract}
 The rapid expansion of Learning Analytics (LA) and Artificial Intelligence in Education (AIED) offers new scalable, data-intensive systems but also raises concerns about data privacy and agency. Excluding stakeholders---like students and teachers---from the design process can potentially lead to mistrust and inadequately aligned tools. Despite a shift towards human-centred design in recent LA and AIED research, there remain gaps in our understanding of the importance of human control, safety, reliability, and trustworthiness in the design and implementation of these systems. We conducted a systematic literature review to explore these concerns and gaps. We analysed 108 papers to provide insights about i) the current state of human-centred LA/AIED research; ii) the extent to which educational stakeholders have contributed to the design process of human-centred LA/AIED systems; iii) the current balance between human control and computer automation of such systems; and iv) the extent to which safety, reliability and trustworthiness have been considered in the literature. Results indicate some consideration of human control in LA/AIED system design, but limited end-user involvement in actual design. Based on these findings, we recommend: 1) carefully balancing stakeholders' involvement in designing and deploying LA/AIED systems throughout all design phases 2) actively involving target end-users, especially students, to delineate the balance between human control and automation, and 3) exploring safety, reliability, and trustworthiness as principles in future human-centred LA/AIED systems.
\end{abstract}
% Keywords
% Each keyword is seperated by \sep


\section{Introduction}

%ai disruption everywhere and in education, 

Advancements in Artificial Intelligence (AI) are rapidly changing how we carry out our daily activities \citep{chakraborty2022artificial}. In educational contexts, AI and learning analytics (LA) innovations are prompting a significant transformation by both enabling innovative teaching and learning strategies \citep{markauskaite2022rethinking} and, at the same time, challenging current assessment practices \citep{swiecki2022assessment}. Emerging LA and AI in Education (AIED) systems offer novel data-intensive solutions that promise to enable adaptive and personalised teaching and learning experiences at scale. For instance, various intelligent tutoring systems and LA solutions provide personalised learning support and automated feedback \citep[e.g.][]{MAIER2022100080,cavalcanti2021automatic}. LA dashboards are providing educators with new means to track student progress and offer targeted support, potentially leading to improved student outcomes \citep[e.g.][]{williamson2022review,fernandez2022beyond}. LA and AIED systems, paired with novel interaction innovations, such as gamification and mixed reality \citep[e.g.][]{carter2021risks}, are also enabling new pedagogical strategies that can potentially make learning more engaging and interactive for students. In sum, educational providers are increasingly adopting LA/AIED systems because data-intensive technologies hold the promise of making learning more accessible, scalable, effective, and personalised for students \citep{macfadyen_institutional_2022} by providing various forms of teacher-facing and student-facing user interfaces \citep{mavrikisSameSameDifferent2021,buckinghamshumLearningAnalyticsAI2019}.

Yet, the proliferation of LA/AIED systems raises key concerns about the privacy of students and security of educational data \citep{viberg2022privacy}, as well as the potential for algorithms to perpetuate biases and discrimination \citep{uttamchandani_introduction_2022}. The lack of involvement of students and teachers in the design and development of LA/AIED systems can potentially lead to a lack of understanding and trust in the technology \citep{shibani2022questioning,dollingerWorkingTogetherLearning2019,sarmiento2022participatory,alzahrani2023teaching}. This lack of involvement also raises questions about the accountability and transparency of AI in education \citep{tsai2020empowering}. Furthermore, increased student adoption of generative AI systems may recurrently challenge existing assessment practices and blur the lines of academic integrity \citep{moya2023academic}.
It is thus essential to address these concerns and incorporate the perspectives and authentic needs of students and teachers in designing and implementing LA/AIED systems. This can potentially contribute to creating ethical, effective, and inclusive ways to adopt LA/AIED systems \citep{williamson2021learning,williamson2022review}.

 In response to the above challenges, there has been a growing interest in adopting human-centred design (HCD) approaches that prioritise human needs, values, and perspectives in the design and deployment of LA/AIED systems \citep{langLearningAnalyticsStakeholder2023,Viberg2023,luckin2006designing,buckingham2019human}. Beyond education, there is also a growing interest in applying similar human-centred principles in the emerging discipline of Human-Centred AI (HCAI). An HCAI approach views AI as systems aimed at serving human interests rather than as a means of achieving technical goals or replacing humans \citep{shneidermanHumanCenteredAI2022a, usmaniHumanCenteredArtificialIntelligence2023}. 
 Thus, the design of HCAI systems is to be guided by a set of ethical principles and design guidelines that empower end-users. The aim is to ensure that technologies providing high automation capabilities are created in ways that guarantee their trustworthiness, transparency, and benefit to society \citep{usmaniHumanCenteredArtificialIntelligence2023,shneiderman2020bridging,ozmen2023six}. 
 
 In recent years, the concept of HCAI has gained significant attention in LA and AIED with a growing recognition of the importance of considering educational stakeholders' needs and perspectives in the design and deployment of data-intensive innovations \citep{buckingham2019human,yangHumancenteredArtificialIntelligence2021,sarmiento2022participatory,holmes2022ethics,kloosH2OLearnHybrid2022, Alfredo2024slade}. This includes, for example, the use of participatory design and co-design methods to involve teachers \citep[e.g.][]{Ahn2021Apr,holstein2019Codesign}, students \citep[e.g.][]{prieto2018co,sarmiento2020engaging}, and other educational stakeholders (such as educational decision-makers, learning designers, IT developers and educational researchers) \citep[e.g.][]{prieto2019orchestrating,schmitz2022fola} in the co-creation of LA/AIED systems, the development of ethical frameworks particularly tailored to LA and AIED \citep[e.g.][]{dollingerWorkingTogetherLearning2019,holmes2022ethics,brossi2022student}, the exploration of new methods for incorporating teachers perspectives and experiences into machine learning algorithms and end-user LA interface designs  \citep{luckin2006designing,wise2021subversive}, and ways to open the AI algorithmic "black box" to provide insight to teachers and students into the analytics or communicate analysis outputs to them in ways they can easily understand \citep{khosravi2022explainable}. 

Given the rapidly growing interest in applying human-centred design principles in LA/AIED research and practice, a timely systematic literature review (SLR) in this area offers a crucial opportunity to synthesise and evaluate the existing body of research. This can inform the direction of future research and ensure that new work is building upon the foundations established by previous studies. 
The closest work to ours was presented by \citet{sarmiento2022participatory}, who conducted an initial, non-systematic review of the literature in participatory and co-design of LA. 
The authors surveyed the design systems and techniques used in the participatory design of learning analytics. 
To the best of our knowledge, no previous research has conducted \textit{comprehensive} and \textit{systematic review} of the body of work focused on applying human-centred design and HCAI principles in LA/AIED literature.

This paper presents a SLR that goes beyond previous works and aims to address the knowledge gap about the state of the art of human-centredness in LA/AIED systems. We used the human-centred AI framework \citep{shneidermanHumanCenteredAI2022a} as a lens to motivate and structure our SLR. This framework helped us to focus on key themes of human-centred design, balance human control and computer automation, safety, reliability, and trustworthiness (SRT), and to identify relevant studies that address these themes in the context of stakeholder involvement in LA and AIED. The findings of this SLR contribute insights into gaps in the existing research, highlighting areas where further investigation is needed and methodological challenges that need to be addressed for human-centred LA and AI in education data-intensive systems to remain relevant and potentially become part of the mainstream practices in the foreseeable future.

 % TODO: Just add one or two to connect LA/AIED/CSCL/EDM. https://www.frontiersin.org/articles/10.3389/feduc.2020.00128/full - https://www.solaresearch.org/2021/12/same-same-but-different-the-fading-boundaries-between-la-and-aied/ - https://discovery.ucl.ac.uk/id/eprint/10110036/1/BJET%202019%20Learning%20Analytics%20and%20AI%20Politics%2C%20Pedagogy%20and%20Practices.pdf
%Similar innovations may come from educational data mining and computer-supported collaborative learning fields, which share boundaries and synergies with LA and AIED, but have distinct focuses, techniques, and educational foundations \citep{rientiesDefiningBoundariesArtificial2020}. 
%\citet{mavrikisSameSameDifferent2021} highlighted the potential for valuable and diverse research in educational technologies by exploring the overlap between the LA and AIED communities.  %Embracing and fostering the interdisciplinary nature of this overlap allows for the development and advancement of inclusive approaches \citep{buckinghamshumLearningAnalyticsAI2019}, leading to significant progress in the field.
\section{Background}

\subsection{Foundations of Human-Centred AI and Learning Analytics in Education}
\label{sec:hcai-definition}

While AI and analytics technologies exhibit potential for augmenting human decision-making, concerns over transparency, accountability, algorithmic bias, discrimination, and potential threats to human autonomy and agency can mitigate their benefits \citep{van2022human}. Consequently, governments -- such as the G20 members \citep{jelinek2021policy} -- and large tech companies \citep[e.g., see review by ][]{jobin2019global} have proposed guidelines for AI application design centred on human values (e.g., safety, reliability, and trustworthiness). In this context, HCAI is emerging as a new research discipline that can be broadly defined as an approach to creating "\textit{software designs that give users high levels of understanding and control over their AI-enabled systems to preserve human agency}" \cite[p.56]{Shneiderman2021issues}. HCAI has its roots in fields such as human-computer interaction, human-centred design, and human factors engineering \citep{chignell2022evolution}, which focus on understanding and incorporating human perspectives and experiences into technology design \citep{grudin2009ai}. HCAI aims to extend human-computer interaction and HCD principles to address unique issues and unforeseen impacts of AI autonomy \citep{xu2023transitioning}.

Echoing \citeauthor{engelbart1962augmenting}'s (\citeyear{engelbart1962augmenting}) seminal research on augmenting intelligence, an HCAI approach emphasises the complementarity of humans and machines, aiming to design AI systems that amplify, augment, and empower individuals by considering SRT principles \citep{shneidermanHumanCenteredAI2022a}. In educational contexts, \citet{buckingham2019human} and \citet{luckin2006designing} were among the pioneers advocating for the integration of human-computer interaction and HCD approaches \citep{giacomin2014human} in the fields of LA and AIED, respectively. They emphasised the need to fully comprehend educational stakeholders' needs, preferences and challenges, often necessitating their inclusion in some stages of the design process by using HCD methods. This reinforces the relevance of HCAI and HCD methods in designing the technical, social, and data-related aspects of LA and AIED, mitigating potential social harms related to the spectrum of analytics use, from rule-based and descriptive to AI and machine learning-driven predictive and prescriptive forms \citep{davenport2018analytics}.

\subsubsection{Human and Computer Automation complementarity. }
\label{sec:hcai-quadrant}
To articulate the balance between humans and AI, \citet{shneidermanHumanCenteredAI2022a} developed a two-dimensional framework. This framework, designed for a broad audience, distinguishes between varying levels of \textbf{\textit{human control}} and \textbf{\textit{computer automation}}. These factors are not considered mutually exclusive. 

% Level of human control
The notion of \textit{\textbf{human control}} is closely related to the \textit{sense of agency}. This subjective experience stems from perceived control over one's actions to make decisions and influence events \citep{mooreWhatSenseAgency2016}. Viewed through the HCD lens, sense of agency is crucial in designing interfaces that support an internal locus of control (i.e., individuals' perceptions of controlling their outcomes).
In an educational setting, human control can range widely, from learners or teachers exercising a high level of agency over the outcomes of LA/AIED systems to simply receiving and understanding information \citep{hooshyarLearningAnalyticsSupporting2023c}. A \textit{high} level of human control may involve learners using their expertise to strategically make informed decisions and take appropriate actions in their learning environment. In contrast, a \textit{low} level of control might occur when individuals receive and understand information but lack autonomy in their learning or teaching environment. This notion of human control encompasses key elements of intentional, purposeful, and meaningful learning \citep{Jaaskela2021Jun} and teaching \citep{Biesta2015Aug}.

% Level in computer automation
The notion of \textbf{\textit{computer automation}} in an HCAI context refers to the characteristics of systems that use computer technology or algorithms to perform tasks automatically, streamlining and expediting operations that were previously completed by human actors (e.g., teachers or students) \citep{Parasuraman2000May}.
The level of automation is determined by its complexity (e.g., ranging from simple rule-based algorithms to complex machine learning and AI models) and the \textit{number} of information process stages the automation supports (i.e., acquiring information, information analysis, decision-making, or action implementation) in the decision-making process \citep{onnaschHumanPerformanceConsequences2014}. In educational setting, a \textit{high} level of automation commonly refers to LA/AIED system capabilities that can automatically capture learner data and make predictions, commonly achieved using multiple algorithms or AI techniques, such as machine learning, computer vision, natural language processing, or intelligent agents \citep[e.g.][]{doganUseArtificialIntelligence2023,jamalovaStateoftheartApplicationsArtificial2022}. On the other hand, a \textit{low} level of computer automation typically involves a system where information is manually prepared or a system follows a fixed set of rules strictly deterministically. Examples can include non-data-intensive solutions (e.g., presentation slide decks) and manually pre-configured visualisations in LA dashboards \citep{fernandez2022beyond}. 
Such systems often lack autonomy and \textit{adaptability} when faced with situations beyond their programmed capabilities.

In turn, a single LA/AIED system can provide various features that combine different levels of human control and computer automation, catering to different tasks and intended users \citep{holsteinConceptualFrameworkHuman2020}. 
This results in the following four quadrants (two-dimensional HCAI framework) contextualised for LA/AIED systems, each representing a different combination of human control and computer automation:
%Representing each dimension into LOW or HIGH will result in 4 quadrants (Q1, Q2, Q3, and Q4): 
\begin{enumerate}
    \label{sec:hcai-quadrant-list}
    \item[Q1:]{\textbf{LOW human control \& LOW computer automation}}. 
    This quadrant represents systems with limited user control or configuration and minimal or no automation. After receiving information from the system, end-users can only comprehend it but cannot personalise or modify it. 
    Some examples in this quadrant include \textit{learning resources} in the learning management system; \textit{student-facing reporting systems} that generate reports for awareness or reflection in an asynchronous manner \citep[e.g.,][]{bodilyTrendsIssuesStudentfacing2017a}; and \textit{rule-based visual data stories} \citep[e.g.,][]{echeverriaExploratoryExplanatoryVisual2018,echeverria2024teamslides,fernandez2024data}, which provide feedback to students about their collaboration process and task completion after a learning activity.
    
    \item[Q2:]{\textbf{HIGH human control \& LOW computer automation}}. 
    This quadrant represents systems where end-users can personalise and configure aspects of the information process and maintain a sense of agency over the learning environment. In contrast, the system facilitates exploration through manual operation with minimal or no automation.
    Examples include \textit{personalised visualisation dashboards} \citep[e.g.,][]{muslimRuleBasedIndicatorDefinition2016}, which allow end-users to control or co-configure the descriptive analytics that learners can use to reflect on the achievement of learning goals; and \textit{educator-driven data analytics systems} \citep[e.g.,][]{fernandez2022beyond}, which may rely on educators' experience, intuition, and informal observations for educators themselves to make decisions about instructional strategies and feedback.
    
    \item[Q3:]{\textbf{LOW human control \& HIGH computer automation}}. 
    This quadrant represents systems that heavily rely on automation with minimal end-user control. Here, automated processes and algorithms handle decision-making and action-taking processes.
    Examples include systems that capture interactions of students or teachers and utilise them in \textit{predictive analytics} components to provide fully \textit{automated feedback} \citep[e.g.,][]{ochoaRAPSystemAutomatic2018, MAIER2022100080} and \textit{automated grading systems} \citep[e.g.,][]{Shetty2022Mar}, which may fully or partly replace the teacher's role in specific assessment tasks.

    \item[Q4:]{\textbf{HIGH human control \& HIGH computer automation}}. 
    This quadrant represents systems that enable manual operation while benefiting from automated assistance to enhance the decision-making process in the learning environment for teachers or learners.
    Examples in this quadrant include \textit{intelligent teaching assistants} that support classroom orchestration \citep[e.g.,][]{lawrenceHowTeachersConceptualise2023}), \textit{recommender systems} that promote the development of metacognitive skills \citep[e.g.,][]{khosraviRiPPLECrowdsourcedAdaptive2019,darvishi2024impact}, and sophisticated modelling/predictive features in \textit{analytics reports or dashboards} that provide guidance and support \citep[e.g.,][]{khachatryanReasoningMindGenie2014}. Here, teachers or students can utilise the highly automated features of the tool to perceive and make sense of learning data and make informed pedagogical decisions about the next course of action \citep{holstein2019Codesign}. 

\end{enumerate}

\subsubsection{The HCAI principles of safety, reliability, and trustworthiness} 
\label{sec:srt-definition}

In addition to defining these levels of control, designing any LA/AIED systems that offer a significant level of human control and computer automation must consider the fundamental principles of safety, reliability, and trustworthiness to ensure ethical practices and reduce the risk of data misuse \citep{cavalcantesiebertMeaningfulHumanControl2023,shneidermanHumanCenteredAI2022a}.
In educational contexts, LA/AIED systems should embrace a \textit{safety} culture, which implies that researchers, designers, and system operators should establish strong relationships with end-users and relevant educational stakeholders, as well as implement strict data privacy measures \citep{holmes2022ethics}. These may include involving stakeholders in data-sharing decisions, ensuring data collection transparency, and granting data access only to specific users \citep{Drachsler2016Apr}. 

An LA/AIED system is deemed \textit{reliable} is when it delivers \textit{accurate} information as anticipated by the user while interacting with it \citep{shneidermanHumanCenteredAI2022a}. However, no predictive model can achieve such perfection \citep{Baker_2016}. 
Therefore, LA/AIED systems designers may benefit from accepting the existence of \textit{imperfection} by studying how users validate and respond to whether such systems may adversely affect their learning \citep{kitto2018embracing}. This can potentially be done by educating humans about AI capabilities and \textit{human biases}, allowing for a balanced delegation of automation tasks while incorporating elements of human oversight \citep{pinskiAIKnowledgeImproving2023}.
Finally, an LA/AIED system is considered \textit{trustworthy} when users have confidence (and, therefore, it is regarded as 'trusted' by users). 
% In other words, users regard the system as 'trusted'. 
Aiming to foster greater trust and confidence among users, the system could actively seek feedback from users, learn from its mistakes, and adapt to improve its performance while aligning with user expectations \citep{usmaniHumanCenteredArtificialIntelligence2023}.
Designers ought to adopt a practical approach that prioritises transparency \citep{Nazaretsky2022Jul,nazaretskyInstrumentMeasuringTeachers2022} and accountability \citep{Pardo2014May} and aim to understand the features of a system that would increase users' trust in it.

\subsubsection{Human-centred design in LA/AIED} 
\label{sec:hcd-background}
An HCAI approach strongly advocates for the participation of pertinent stakeholders, such as end-users with \textit{lived experience}, potential future users, policymakers, and experts in ethics and human values, in the design and deployment of AI systems \citep{usmaniHumanCenteredArtificialIntelligence2023,shneidermanHCAI2020}. 
HCD thus plays a crucial role in HCAI as it focuses on the needs and requirements of the people for whom the system is intended, rather than the designer's creative process or the technology capabilities \citep{giacomin2014human}. 
As a participatory practice, HCD involves an iterative process of understanding the context, identifying end-user requirements, and involving stakeholders with lived experiences in the design and evaluation of the system, ideally engaging with stakeholders as equal partners \citep{xu2023transitioning}. From \citeauthor{haningtonMartin2012}'s (\citeyear{haningtonMartin2012}) perspective, the HCD process includes five multifaceted phases: (1) \textit{planning, scoping, and definition} -- clearly defining the parameters of the project; (2) \textit{exploration, synthesis, and design implications} -- conducting immersive research and design ethnography to gather information and derive insights that will guide design choices; (3) \textit{concept generation and early prototype iteration} -- generating ideas, concepts, or creating an early version of prototypes with stakeholders; (4) \textit{evaluation, refinement, and production} -- testing and gathering feedback from stakeholders to refine designs, ensuring they meet the desired standards in the production; and (5) \textit{launch and monitor} -- conducting quality assurance testing of the design to ensure readiness for public use, and ongoing analysis to make necessary adjustments if needed.

Numerous HCD techniques are available to assist designers and researchers in the aforementioned phases \citep{giacomin2014human, maguireMethodsSupportHumancentred2001}. These techniques can enable interaction with stakeholders and aid in the identification of their meanings, desires, and needs, which can be achieved through verbal techniques (e.g., ethnographic interviews, questionnaires, cognitive tasks, think-aloud, persona crafting, and brainstorming) or non-verbal techniques (e.g., probing, observations, body language analysis, and physiological analysis). A growing set of more speculative techniques (e.g., real fiction, role-playing, para-functional prototypes, what-if scenarios and fabulation) are used for simulating opportunities and envisaging possible futures for designing future-looking features.
Nowadays, some of these HCD techniques are starting to be used to design contemporary LA/AIED systems or to envisage potential future scenarios of AI application in education \citep[e.g.,][]{prestigiacomo2020learning,holstein2019Codesign,echeverriaExploringHumanAI2020}.

Regarding stakeholder involvement, \citet{sarmiento2022participatory} conducted an initial review of HCD methods in the LA literature, particularly in participatory design and co-design. They pointed out the increased use of these two methods in higher education and highlighted detailed descriptions of research techniques that were often lacking. 
From this point of view, the extent of stakeholders' involvement can be categorised into \textit{active} or \textit{passive}. \textbf{Active involvement} refers to possessing agency in shaping the outcomes that arise from the design processes \citep{dollingerWorkingTogetherLearning2019}. Stakeholders engage in conversations that contribute to creating designs, assisting in testing, evaluating, and providing feedback on designs, or being actively involved in decision-making processes that shape the design of LA/AIED systems. 
On the other hand, \textbf{passive involvement} refers to stakeholders playing a role as more consultative or advisory in the design activities \citep{Edelenbos2006Jul}. They provide input, feedback, or opinions but may have limited decision-making authority or direct influence over the outcomes. 
% ---- END OF GENERAL BACKGROUND ---

% The next section provides an overview of these research works.  
\subsection{Related Works and Research Gaps}
\label{sec:related-work}
To better understand how human-centredness has been considered in the design of LA/AIED systems, it is crucial to examine the key \textit{characteristics} of current research in this area. These include the education levels mainly targeted, the types of research methods used (such as qualitative, quantitative, or mixed), and the most commonly applied HCD techniques like co-design, focus groups, and prototyping.
For instance, previous reviews of LA/AIED generally have focused on higher education \citep[e.g.,][]{williamson2022review, leitnerLearningAnalyticsHigher2017, doganUseArtificialIntelligence2023}, K-12 \citep{zhangAITechnologiesEducation2021, Granic2022Aug, linArtificialIntelligenceIntelligent2023}, or informal learning \citep{Granic2022Aug}. However, it remains unclear which education level has most thoroughly considered human-centredness. Moreover, there is a gap in current research regarding the diverse research methodologies used in human-centred LA/AIED research and the specific design phases (listed in Section \ref{sec:hcd-background}) where they have been applied. At the same time, previous non-systematic reviews that initially explored human-centredness in LA \citep[e.g.,][]{sarmiento2022participatory} and AIED \citep[e.g.,][]{khosravi2022explainable} literature have not identified the specific HCD techniques employed at different design stages. 
Together, these issues motivate our first research question:
    \begin{quote}
     \textbf{RQ1.} What is the current state of human-centred LA/AIED research, specifically through the lens of education levels targeted, research methodologies employed, design phases covered, and HCD techniques utilised?
    \end{quote}

Arguably \citep{langLearningAnalyticsStakeholder2023}, the involvement of educational stakeholders is paramount in the human-centred design process, as highlighted in recent human-centred LA \citep[e.g.,][]{sarmiento2022participatory, barreirosStudentsFocusMoving2023} and AIED \citep[e.g.,][]{liRiskFrameworkHumancentered2023, linArtificialIntelligenceIntelligent2023} works. However, this involvement can take various forms \citep{langLearningAnalyticsStakeholder2023}. It remains unclear to what extent educational stakeholders' involvement, whether \textit{active} or \textit{passive} (see Section \ref{sec:hcd-background}), has been considered in the various phases of the design process, from inception to implementation and evaluation of existing human-centred LA/AIED systems. Additionally, none of these works have discussed how extensively the perspectives of each educational stakeholder (i.e., teachers, students, and experts) have been incorporated at the various design phases. This highlights a pressing need to explore the extent of each educational stakeholder's contribution to the design process of current human-centred LA/AIED systems. This motivates our second research question:
    \begin{quote}
        \textbf{RQ2:} To what extent have educational stakeholders contributed to the design process (phases) of current human-centred LA/AIED systems?
    \end{quote}

Although the involvement of stakeholders in the design of LA/AIED systems has received significant attention \citep[e.g.,][]{lawrenceHowTeachersConceptualise2023,holstein2019Codesign,kaliisaChecklistGuidePlanning2023} other key aspects of human-centredness, such as balancing system features for end-users against fully automated features focused on technical goals or replacing human activity, are also important \citep{langLearningAnalyticsStakeholder2023, usmaniHumanCenteredArtificialIntelligence2023}. 
LA/AIED researchers have suggested various ways to empower users, such as offering control over automated recommendations \citep[e.g.,][]{maGlanceeAdaptableSystem2022,lawrenceHowTeachersConceptualise2023} or enhancing the explainability of AI outcomes \citep[e.g.,][]{khosravi2022explainable,kloosH2OLearnHybrid2022}.
However, the balance between human control and computer automation in existing human-centred LA/AIED systems remains underexplored. 
Additionally, the importance of empowering stakeholders in light of rapid AI advancements was emphasised in a systematic review by \citet{doganUseArtificialIntelligence2023}, which focused on the implementation of AI in online learning but not in other LA/AIED system types like visualisation/dashboards, intelligent tutoring systems, and recommender systems \citep{kaliisaChecklistGuidePlanning2023, williamson2022review, linArtificialIntelligenceIntelligent2023, daSilva2023Mar}. To our knowledge, no studies have examined the level of human control and computer automation in designing existing human-centred LA/AIED systems.
The two-dimensional HCAI framework proposed by \citet{shneidermanHCAI2020} (see Section \ref{sec:hcai-quadrant-list}) provides a frame of reference to analyse the human control/automation balance. 
This motivates our third research question:
\begin{quote}
    \textbf{RQ3:} What levels of human control and computer automation have been considered in various types of human-centred LA/AIED systems designed with the involvement of stakeholders? 
\end{quote}

% --- RQ4
As mentioned above, some researchers have increasingly recognised the importance and potential of HCAI in the current LA/AIED literature landscape \citep[e.g.,][]{renzReinvigoratingDiscourseHumanCentered2021,zhaoLearningAnalyticsFramework2023}. One critical aspect that has garnered considerable attention within the HCAI domain is the assurance of safety, reliability, and trustworthiness in such systems.
For example, \citet{renzReinvigoratingDiscourseHumanCentered2021} highlighted the importance of incorporating human values and ethical considerations while developing AI systems for personalised learning environments. The authors visualised the applications of AIED systems \citep[fig. 4, p.12]{renzReinvigoratingDiscourseHumanCentered2021} in an adapted \citeauthor{shneidermanHCAI2020}'s (\citeyear{shneidermanHCAI2020}) two-dimensional HCAI framework, by adding a \textit{trustworthiness dimension} perpendicular to human-AI augmentation and types of machine learning. 
In a more recent LA study, \citet{zhaoLearningAnalyticsFramework2023} proposed a framework based on HCAI to identify the most effective learning strategies highlighting the significance of \textit{reliability} of the AI algorithms. 
In AIED literature, \citet{yangHumancenteredArtificialIntelligence2021} argued that AI systems should be designed to be transparent, explainable, and accountable to reduce the risk of algorithmic bias and misuse of AI. 
While these examples recognise the critical role of investigating \textit{safety, reliability, and trustworthiness} principles, there is a lack of understanding of how these principles have been considered when designing human-centred LA/AIED systems. 
This motivates our fourth research question:
\begin{quote}
        \textbf{RQ4:} How and to what extent have the HCAI principles of safety, reliability, and trustworthiness been discussed in existing human-centred LA/AIED systems?
\end{quote}

\section{Method}

\subsection{Review Procedures}
\begin{figure*}[h!]
    \centering
    \includegraphics[width=0.8\textwidth]{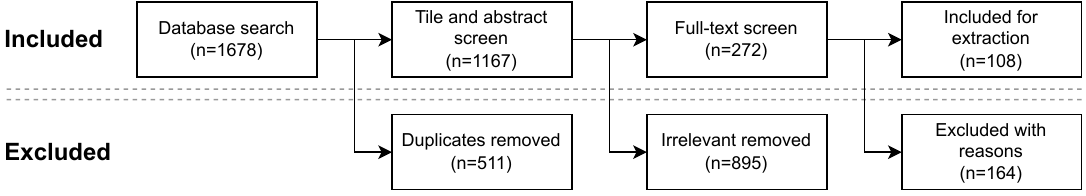}
    \caption{The PRISMA protocol as applied in the current systematic literature review}
    \label{fig:prisma}
\end{figure*}

To conduct the systematic literature review, we followed the Preferred Reporting Items for Systematic Reviews and Meta-Analyses (PRISMA) protocol \citep{Page2021Mar}, which has four phases and aims to promote transparent reporting. We searched four reputable bibliographic databases, including Scopus, ACM Digital Library, IEEE Xplore, and Web of Science, to find high-quality peer-reviewed publications on HCAI in LA, and where the relevant LA/AIED (e.g., \textit{Learning Analytics and Knowledge} -- LAK and \textit{Artificial Intelligence in Education} -- AIED) conferences are commonly published (i.e., ACM and Scopus, respectively). The initial query that was used for the title, abstract, and keyword search of peer-reviewed publications included the following groups of keywords: 
\begin{itemize}
    \item \textit{(human-cent*red OR user-cent*redness OR "value-sensitive design" OR "co-design" OR "participatory design" OR "co-creation" OR design process) AND 
    \item ("learning analytics" OR education* OR student* OR teacher* OR classroom) AND 
    \item ("artificial intelligence" OR "intelligent augmentation" OR tool* OR system OR automation OR AI OR analytic* OR algorithm* OR visuali*ation OR "dashboard" OR "interface")}. 
\end{itemize}
To focus on studies published in the established research field of LA and AIED in the last decade, a publication year constraint was used to search for relevant publications from January 1, 2012 to November 1, 2023.
By focusing on the past 10 years, this SLR captures the most recent advancements in LA/AI technologies and their educational applications, providing an up-to-date understanding of the field's current state.
The database search initially yielded 1,678 articles with 511 duplicates removed (14 duplicates were identified and merged manually), leaving 1167 articles for the title and abstract screening process (see Figure \ref{fig:prisma}).
To ensure a thorough and accurate screening, three researchers independently reviewed the titles and abstracts of eligible articles according to three predetermined inclusion and exclusion criteria, as follows: 
\begin{enumerate}
  \item First, we included articles that reported empirical studies for developing LA/AIED systems (long papers). These included studies proposing a design framework that may use tools/interfaces for illustration/explanation in practice. We \textit{excluded} non-empirical studies, such as theoretical, opinion/positioning, dataset, literature review, and short papers.
  \item Second, we included articles that reported on the studies that aimed to design or develop data-intensive education systems/tools/interfaces for end-users and excluded studies that only mentioned “education”, “learning", "LA", or "AI" as an example of other more general topics.
  \item  Third, we only \textit{included} articles that involved stakeholders (e.g., teachers, students, administrative staff, and designers) in the study and \textit{excluded} studies that focus merely on advancing technical aspects of the technology (e.g., improving the accuracy of AI algorithms) without directly studying the implications for human learning.
\end{enumerate}
At least two researchers reviewed each article (R1-R2), and a third researcher (R3) resolved the conflicts through in-depth simultaneous discussions until a consensus was reached. 
After the title and abstract screening, a total of 272 articles (as seen in Figure \ref{fig:prisma}) were identified as candidates for full-text review. The inter-rater reliability among the three researchers was 0.80 (R1\&R2), 0.88 (R1\&R3), and 0.74 (R2\&R3), as measured by Cohen's kappa, indicating a substantial to high agreement among the reviewers \citep{Page2021Mar}. 

After the full-text review, 164 articles were excluded following our exclusion criteria (detailed above), consisting of \textit{not aiming to design or develop LA/AIED systems for end-users} (n=81), \textit{non-empirical study} (n=53), \textit{merely mentioned "education"/"learning"/"AI"/"LA" as examples} (n=15), \textit{focused on advancing algorithms/technologies without learning implications} (n=9), \textit{lack of full-text access} (n=6).  
This means that 108 articles were chosen for data extraction, and the inter-rater reliability (Cohen's kappa) for the full-text screening was 0.71 (R1\&R2) and 0.67 (R1\&R3), indicating a substantial agreement between the researchers. Conflicting decisions were resolved through in-depth discussions among researchers (R1--R3) or consulting a fourth researcher to reach a consensus \citep{Yan2022Mar}.
The following subsections describe the data extraction and analysis processes to address each research question.

\subsection{Data extraction and analysis}
\label{sec:analysis}

%RQ1
To address \textbf{RQ1}, we collected the study characteristics, including targeted educational levels (i.e., K-12, middle school, high school, and university/college) and employed research methodologies (i.e., qualitative, quantitative, or mixed).
Next, we employed the five multifaceted design phases by \citet[][see Section \ref{sec:hcd-background}]{haningtonMartin2012} to identify in which phase(s) stakeholders' views or inputs were considered in the studies included in the review.
We also collected information about what HCD techniques were utilised in the study (e.g., co-design, prototyping, interviews, and storyboards). It is important to note that each study could employ multiple HCD research techniques throughout different design phases. We presented our findings in the form of a Sankey Diagram to inductively analyse relationships between characteristics, design phases, and HCD techniques.

%RQ2
To address \textbf{RQ2}, we extracted the information on stakeholders' involvement as active or passive, aligning with their definitions presented above (in Section \ref{sec:hcd-background}). We further identified the level of involvement based on stakeholders' roles, such as students, teachers, subject experts, and administrators. 
To ascertain the impact of stakeholder involvement on the design of the LA/AIED system, we mapped their active or passive involvement status in each of five multifaceted design phases proposed by \citet[][see Section \ref{sec:hcd-background}]{haningtonMartin2012}. For example, \citet{holstein2019Codesign} actively involved teachers in developing a classroom orchestration system for their use, representing active involvement. On the other hand, \citet{fernandez2022beyond} consulted teachers in designing a system for student use, in which students who would be the intended users were not involved in any of the phases of the design process. 
All this information was subjected to quantitative analysis by comparing and employing descriptive statistics as the primary analytical approach for reporting results. The percentages presented in the next section were calculated based on the total of the included studies (n=108).

%RQ3
To address \textbf{RQ3}, first, the proposed human-centred LA/AIED system described in each article was categorised based on the definition of the two-dimensional HCAI framework \citep{shneidermanHumanCenteredAI2022a}, which included the four quadrants of human control and computer automation outlined in Section \ref{sec:hcai-quadrant-list}.
Then, we further mapped these systems with the aforementioned stakeholder involvement (active or passive, see \ref{sec:hcd-background}).
We also collected the type of LA/AIED system (i.e., visualisation/dashboard, intelligent tutoring system, learning design tool, recommender system, prediction system, information retrieval assistant, and evaluation of course essay system) presented in each article \citep{zhangAITechnologiesEducation2021,Granic2022Aug,daSilva2023Mar}. 
This gathered information underwent inductive analysis through a comparative approach, incorporating descriptive statistics and supported by illustrative examples.

% RQ4
Lastly, regarding \textbf{RQ4}, any discussion about HCAI principles within the scope of safety, reliability and trustworthiness was extracted with a question \textit{"How did authors discuss safety, reliability, or trustworthiness?"}. We qualitatively analysed these discussions using inductive thematic analysis \citep{braun2012thematic}.
We prepared and applied an initial coding scheme that pertained to safety, reliability and trustworthiness (as defined in Section \ref{sec:srt-definition}). In this case, data \textit{safety} included the subcodes: data privacy, data sharing, and transparency in data collection. \textit{Reliability} encompassed the subcodes: data accuracy, completeness, bias, validity, and consistency. \textit{Trustworthiness} included the subcodes: trust, transparency, and accountability.
The coded data was cross-checked between two researchers concurrently, and any emerging codes or conflicts were jointly agreed upon in iterative discussion sessions until consensus was reached. 
Lastly, the emerging themes were reported by classifying them into the context of safety, reliability and trustworthiness.

\begin{comment}

we determined whether the stakeholders considered in the study were \textit{direct stakeholders} (those who can interact with and have control over the system) or \textit{indirect stakeholders} (those who may not interact with the system but are likely to be impacted by tool usage), based on the stakeholder analysis classification proposed by \citet{chenValueSensitiveLearningAnalytics2019}.
We further mapped these with the aforementioned levels of stakeholder involvement (either active or passive, see \ref{sec:hcd-background}).

%This approach enabled us to gather empirical data from existing literature and draw inductive conclusions on key characteristics of stakeholder involvement in designing the LA/AIED system.
%Next, our initial code scheme coded the extracted data from the papers. 
and merging studies that were investigating the same project (n=16)
    %This constraint was also based on the formal establishment of the LA research field after 2010, coinciding with the launch of the LAK conference in 2011. 
    % the learning contexts (i.e., individual learning, group learning, mentoring, professional development, and informal learning)
    % Additionally, we extracted the descriptions of human control and computer automation features, including information regarding target end-users who will control the tool. 
%As stated by \citet{McDonald2022}, the need for inter-rater reliability in this process was not necessary considering our decision-making process involved multiple iterations and simultaneous consensus

\end{comment}
\section{Results}

\subsection{Current research on  human-centred LA/AIED systems (RQ1)}
% The most significant portion of studies came from the United States, accounting for 28\% of the total studies. Countries in Europe (e.g., Germany, Switzerland, Netherlands, Sweden) consisted of 28\% of the studies, followed by Australia with 13\%, UK with 8\%, and China with 3\%. Other countries have smaller percentages, including Canada, Singapore, Brazil, Japan, Pakistan, Bahrain, Ecuador, and Colombia. These countries collectively made up around 21\% of the studies. The remaining countries had a lower representation, accounting for approximately 1\%. 

Most studies were reported at a University/college level (53\%), followed by K-12 (34\%) and informal learning (13\%).
% research methods
The type of research method reported in these studies was distributed as follows: qualitative (44\%), mixed-methods (43\%), and quantitative (13\%).
Regarding the design phases of the HCD process, we found that 18\% of the studies addressed Phase 1 (planning, scoping, and definition). Phase 2 (exploration, synthesis, design implications) was addressed in 53\% of the studies, while Phase 3 (concept generation and early prototype iteration) accounted for 51\%. Phase 4 (evaluation, refinement, and production) received the highest attention at 56\%. However, only a small minority, 9\% of the studies, considered Phase 5 (launch and monitor).

Next, common HCD techniques used in the design process include interviews (52\%), questionnaires (47\%), co-design (28\%), prototype validation (26\%), focus groups (10\%), observations (10\%), and surveys (8\%). Less used techniques included workshops (5\%), storyboards (4\%), personas (4\%), and card sorting (3\%). Very few studies used other techniques that are common in wider HCD research, such as Wizard-of-Oz  \citep{schulzTutoringSystemSupport2022,vinellaFormingTeamsLearners2022,echeverriaExploringHumanAI2020}, speed dating \citep{holstein2019Codesign,tenorioExploringDesignConcepts2022}, think-aloud \citep{conijnHumancenteredDesignDashboard2020, Ahn2021Apr}, eye tracking \citep{lalleImpactStudentIndividual2017, mangaroskaGazeInsightsDebugging2018}, and user journeys \citep{weigandHCD3AHCDModel2021}. 

\begin{figure*}[h!]
    \centering
    \includegraphics[width=0.90\textwidth]{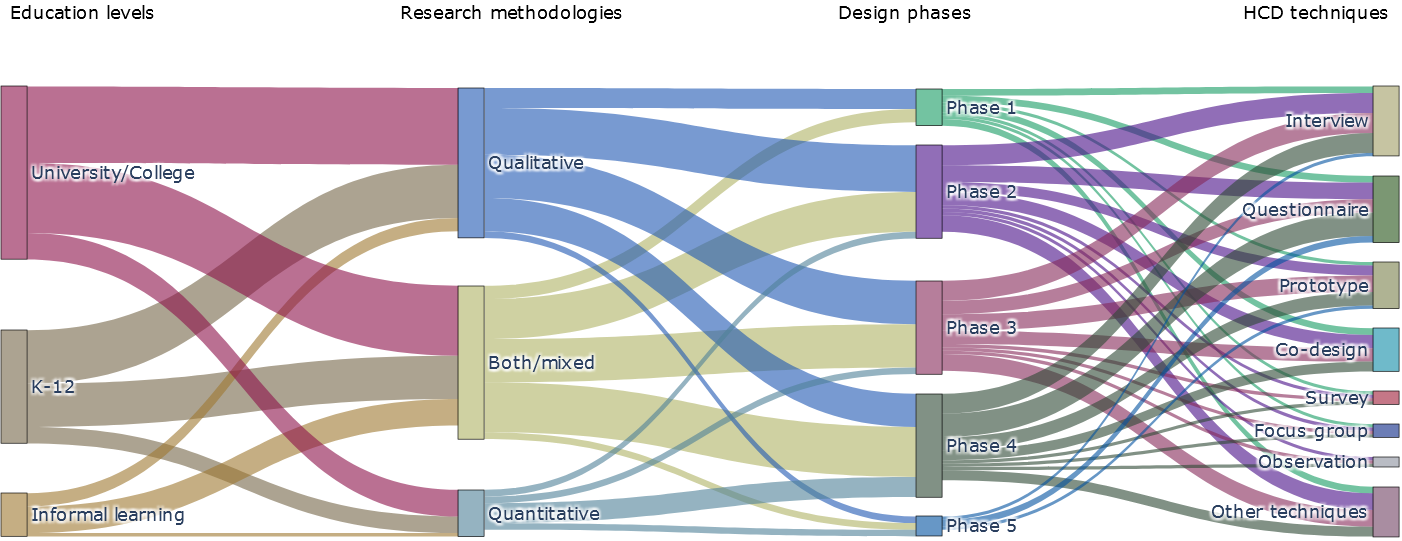}%
    \caption{The distribution of key characteristics in the current research state of human-centred LA/AIED systems, depicted through a Sankey diagram. Categories are sorted by frequency.
    The diagram illustrates the distribution of education levels, the utilisation of research methodologies (qualitative, quantitative, or mixed), the progression through five multifaceted design phases, and the application of HCD techniques. The varying thickness of the flow path represents the significance of each transition. 
    }
    \label{fig:rq1-sankey}
\end{figure*}

Figure \ref{fig:rq1-sankey} illustrates the characteristics of current research focused on human-centred LA/AIED systems. In summary, we found that most studies have been conducted at the university level, and studies have mostly involved qualitative or mixed-method research. More attention has been given to Phases 2, 3, and 4 of the HCD process, while Phases 1 and 5 received less attention. There was a shift from a more qualitative methodology approach in Phase 2 to a more mixed methodology in Phase 4 as the design tended to mature. Quantitative research has not been employed in Phase 1. Yet, some quantitative methods are used at later stages of the design process (phases 4 and 5), particularly in the form of evaluation questionnaires. Last but not least, interviews have been the most commonly used HCD technique (20\%) in most design phases, except for Phase 4 (5\%), where questionnaires were more prevalent (7\%).

\subsection{The extent of stakeholder involvement in design (RQ2)}
% RQ2: What is the current state of learning analytics systems that follow HCD, specifically regarding the level of education, stakeholders, process, and techniques? 

\begin{figure*}[h!]
    \centering
    \includegraphics[width=0.8\textwidth]{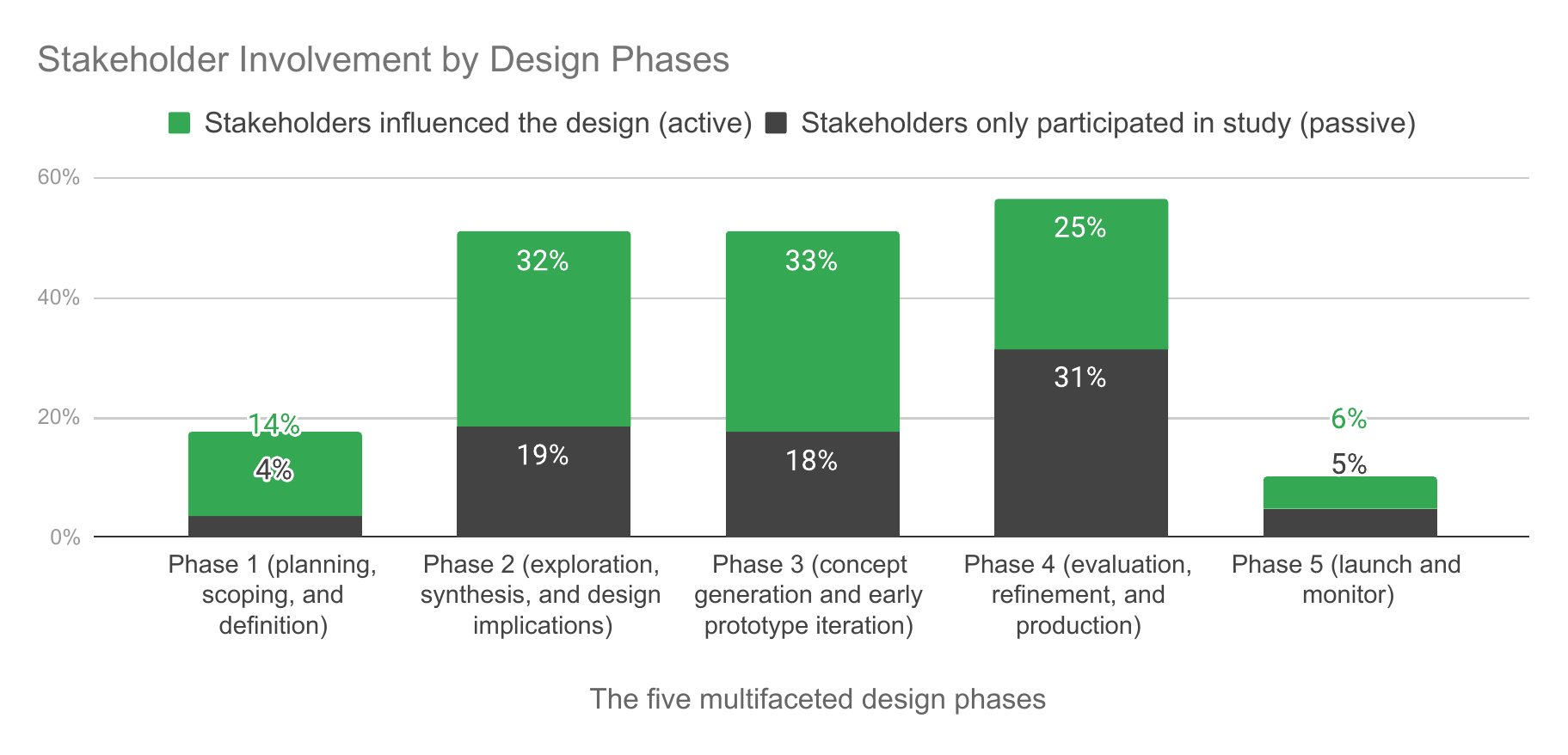}%
    \caption{Distribution of stakeholder involvement varies across various design phases. The data demonstrates shifts in passive (black) and active (green) involvement throughout design phases, with more passive stakeholders in Phase 4.}
    \label{fig:rq1-design-phases}
\end{figure*}

Concerning stakeholder involvement in the design process, stakeholders were classified into active and passive roles (see definitions in Section \ref{sec:hcd-background}). 
Figure \ref{fig:rq1-design-phases} depicts each design phase with the percentage of included studies covering stakeholders' active/passive involvement. 

Overall, lower (both active and passive together) stakeholder involvement has been observed in Phases 1 (18\% -- 4\% passive and 14\% active) and 5 (10\% -- 5\% passive and 6\% active), while there has been more stakeholder involvement in other design phases (Phase 2 -- 51\%, Phase 3 -- 51\%, and Phase 4 -- 56\%).
Our findings also indicated a higher \textbf{active} stakeholder involvement in Phase 2 (32\%) and Phase 3 (33\%) compared to the \textbf{passive} involvement. 
%PHASE 4
Notably, the design \textbf{Phase 4} -- evaluation, refinement, and production -- was the only phase that had higher passive (31\%) compared to active (25\%) stakeholder involvement. This finding suggests that while most of the works are concentrated on evaluating the design of their LA/AIED systems, there is a tendency for the outcomes produced by stakeholders not to be further considered for improving these designs. 
For example, \citeauthor{khosraviRiPPLECrowdsourcedAdaptive2019}'s (\citeyear{khosraviRiPPLECrowdsourcedAdaptive2019}) study illustrates this \textit{passive} stakeholder involvement, where the authors only reported on students' evaluation of a system through surveys and lab experiments. 
% Yet, researchers made final decisions from the statistical conclusions to refine the system.
In contrast, \citeauthor{wileyHumancentredLearningAnalytics2023a}'s (\citeyear{wileyHumancentredLearningAnalytics2023a}) work illustrates \textit{active} involvement in which teachers participated in several iterative evaluation sessions to re-design an LA dashboard based on their continuous input and feedback.

\begin{figure*}[h!]
    \centering
    \includegraphics[width=0.9\textwidth]{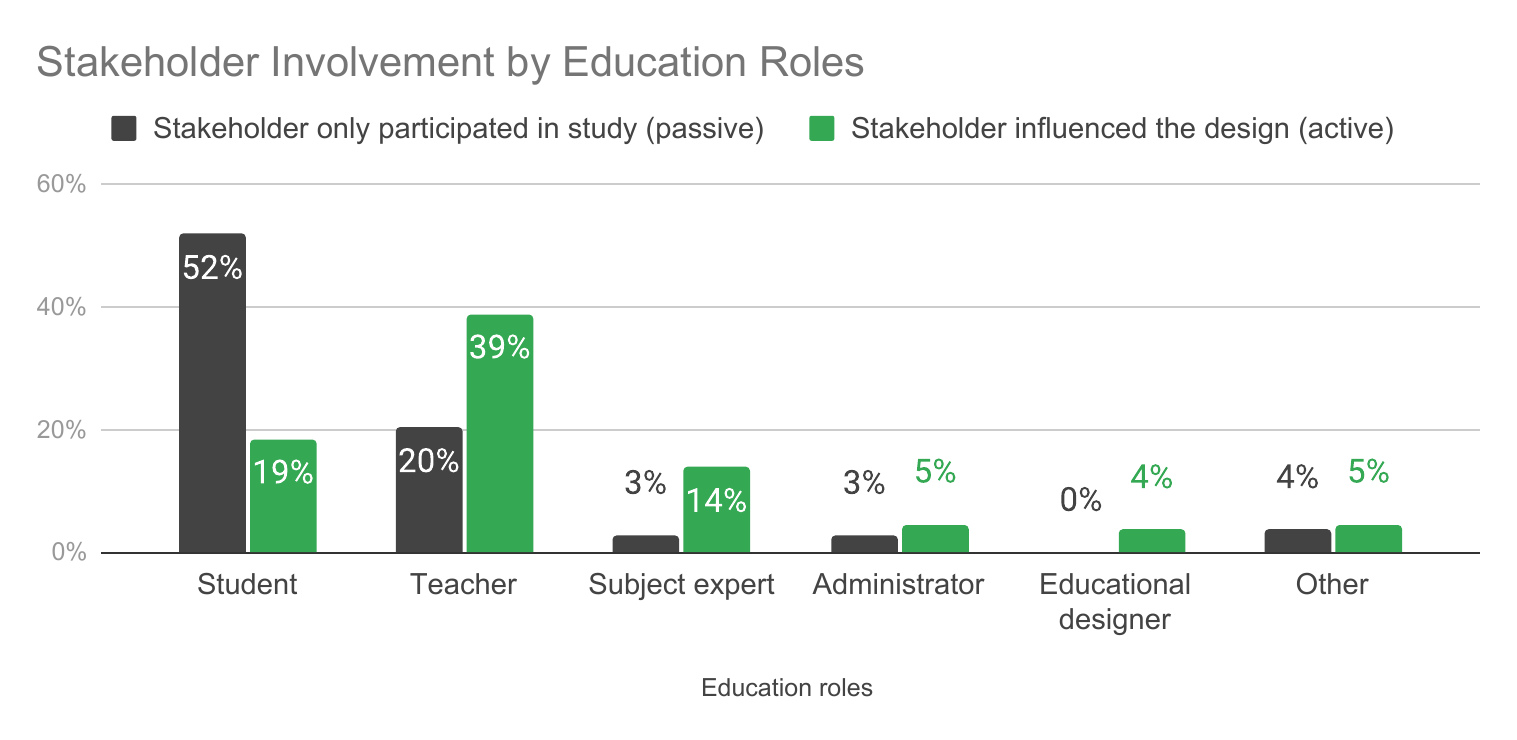}%
    \caption{Distribution of stakeholder involvement in the design of human-centred LA/AIED systems. The green bar represents the percentage of stakeholders actively contributing to and influencing the design. In contrast, the black bar represents the percentage of stakeholders participating in the study with limited influence on design outcomes.}
    \label{fig:rq1-stakeholders-involvement}
\end{figure*}

% EDUCATION STAKEHOLDERS BY ROLE
Furthermore, our analysis revealed that students and teachers were involved in 71\% and 59\%, respectively of the total number of included studies (see Figure \ref{fig:rq1-stakeholders-involvement}). Subject experts made up 17\%, administrators 8\%, and educational designers 4\%. The remaining 9\% included professionals such as software developers \citep[e.g.,][]{wileyHumancentredLearningAnalytics2023a}, evaluators \citep[e.g.,][]{ocumpaughGuidanceCounselorReports2017}, and counsellors \citep[e.g.,][]{cukurovaInteractionAnalysisOnline2017}.

% Regarding the degree of stakeholder involvement across different education roles, 
Students exhibited the highest representation of passive involvement at 52\%.
For instance, students are often involved in the evaluation of the system \citep[e.g.,][]{ochoaControlledEvaluationMultimodal2020,limImpactLearningAnalytics2021}, or being observed on their behaviours during a learning activity while using the system \citep[e.g.,][]{khosraviRiPPLECrowdsourcedAdaptive2019} but this does not necessarily means they are actively involved in the design of such systems. 
The active student involvement can be observed, for example, in the form of participating in ideas generation design activities \citep[e.g.,][]{dequinceyStudentCentredDesign2019,wangCoDesigningAIAgents2022}. 
Conversely, despite students being the most involved stakeholders, their active involvement was comparatively lower than that of teachers (39\%). 
In these studies, teachers were involved in co-design activities, from inception to evaluation, where the systems were designed to meet their teaching needs and preferences \citep[e.g.,][]{lawrenceHowTeachersConceptualise2023, rodriguez-trianaAdaIblLessons2021, olsenDesigningCoOrchestrationSocial2021}. 
An example of passive involvement of teachers (20\%) is when they are invited to explore the system, thereby offering researchers valuable insights into the sensemaking process without clarifying whether such insights are further considered in the design or re-design of the system \citep[e.g.,][]{fernandez-nietoClassroomDandelionsVisualising2022, alfredoThatStudentShould2023a}.

Subject experts \citep[e.g.,][]{eradzeContextualisingLearningAnalytics2020} and educational designers \citep[e.g.,][]{tsaiChartingDesignNeeds2022}, 14\% and 4\% respectively, were more actively involved through participatory design. Administrators and others had almost equal proportions of active roles at 5\% \citep[e.g.,][]{desilvaInstitutionalAnalyticsAgenda2022,bonnatDigitalCompanionMonitor2022a}.
In summary, students and teachers were the most involved education stakeholders, with students having the highest representation of passive involvement and teachers having the highest representation of active involvement.

% RQ3: What are the human control and AI automation levels in current learning analytics systems?
\subsection{The levels of human control and computer automation (RQ3)}

We examined the distribution of LA/AIED systems across the four quadrants of the HCAI framework (see Section \ref{sec:hcai-quadrant}). 
% These systems were categorised into Q1 (12\%, n=12, \citep{}), Q2 (31\%, n=33, \citep{}), Q3 (11\%, n=11, \citep{}), and Q4 (47\%, n=50, \citep{}). 
% NOTE[good to add at the first beginning]: Overall perspective, mostly are active, except Phase 4. After this paragraph. Update first paragraph. Active more prevalent vs passive prevalent (separate paragraph).
This distribution is summarised in Figure \ref{fig:rq2-hcai} and the details are in the Appendix (see Table \ref{table:appendix-reference}).
\begin{figure*}[ht]
    \centering
    \includegraphics[width=0.75\textwidth]{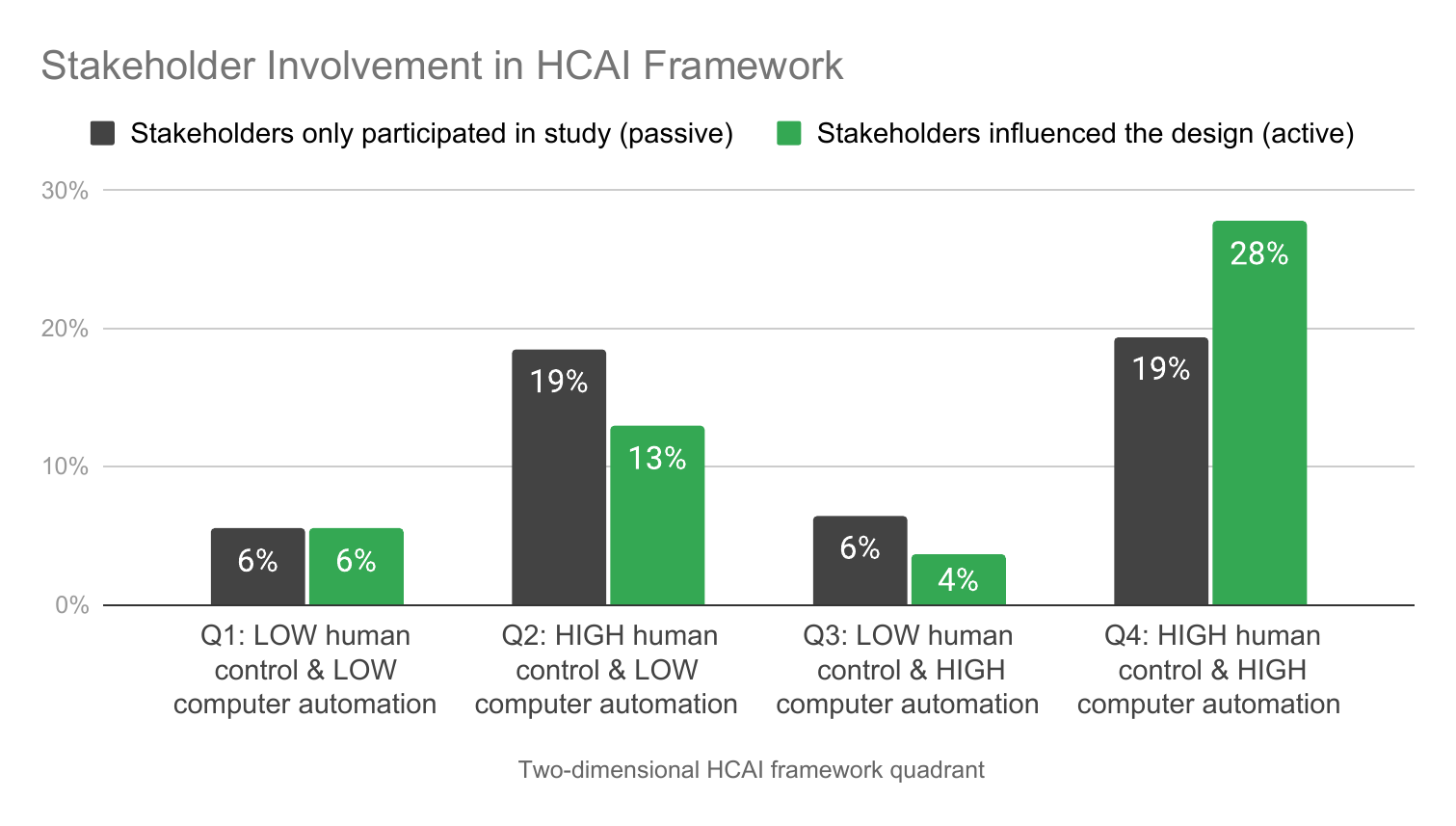}%
    \caption{Distribution of human-centred LA/AIED systems categorised into the human control and computer automation quadrants of the HCAI framework (Q1-Q4, see definitions in Section \ref{sec:hcai-quadrant}). The green bars represent studies where the stakeholders actively engaged and contributed to the design of the tools. The black bars represent studies where stakeholders merely participated as test subjects, offering limited input during the design phase of LA/AIED systems. 
    }
    \label{fig:rq2-hcai}
\end{figure*}

Notably, a higher proportion of human-centred LA/AIED systems have already considered substantial human control features in their design, with a more significant percentage of systems categorised in Q4 at 47\% \citep[e.g.,][]{pozdniakovQuestiondrivenDashboardHow2022,alfredoThatStudentShould2023a} and in Q2 at 32\% \citep[e.g.,][]{huITalkISeeParticipatory2022a,jeongAutomataStageARMediatedCreativity2023a}. In contrast, a smaller proportion is found in Q1 at 12\% \citep[e.g.,][]{kivimakiCurricularConceptMaps2019, fernandez-nietoModellingSpatialBehaviours2021} and in Q3 at 10\% \citep[e.g.,][]{ochoaRAPSystemAutomatic2018, ocumpaughGuidanceCounselorReports2017}. 
Next, we present the findings of Q4 and Q2, as these were the most frequent categories, followed by Q1 and Q3, which were the least represented categories. 

% Q4
LA/AIED systems in \textbf{Q4} have had more active stakeholder involvement (28\%) compared to systems in other quadrants, where stakeholders have often been passively involved.
One example for Q4 is illustrated in \citet{lawrenceHowTeachersConceptualise2023}'s work, in which teachers were actively involved in several design phases during a multi-year study to build an AI-powered classroom orchestration system that explicitly considers teachers' agency (i.e., giving the control to teachers to accept or reject recommendations coming from the AI).
In contrast, passive stakeholder involvement in Q4 (19\%) include studies where researchers documented observations and behavioural multimodal data (eye tracking data, interaction log files, and other physiological sensors) from students interacting with an AI-powered system (i.e., intelligent tutoring system) \citep[e.g.,][]{lalleImpactStudentIndividual2017}.
% e.g., \citep{liaqatCollaboratingMatureEnglish2021,lawrenceHowTeachersConceptualise2023}

% Q2
Next, systems in \textbf{Q2} had less active stakeholder involvement in their design processes (13\%) compared to a higher passive involvement (19\%). In this quadrant, active stakeholder involvement includes participatory design activities to understand how stakeholders use systems that require customisation (i.e., manually selecting options to generate visualisations) on exploratory tasks, for example, how teachers use a data visualisation inquiry tool that requires them to control and customise visualisations \citep[e.g.,][]{shreinerInformationWonJust2022a}. In contrast, passive involvement may be observed in the form of usability studies where participants are asked to evaluate systems through surveys or questionnaires, such as in the work by  \citet{muslimRuleBasedIndicatorDefinition2016}, where students and teachers were asked to rate the usefulness of a rule-based system to support flexible definition and dynamic generation of indicators to meet the needs of different stakeholders with diverse goals and questions (i.e., exploratory tasks using LA dashboards). This can be considered an indirect influence on design outcomes. 

% Q3
Regarding systems in  \textbf{Q3} which feature high automation and low human control, we found less active stakeholder involvement in their design process (4\%) compared to passive involvement (6\%). For Q3, active involvement can be illustrated through co-design sessions aimed at identifying features that could be included in intelligent systems, for example, by including teachers' perspectives in the design of two modalities of a robot to support social interactions (as a social actor) or knowledge acquisition (as a didactic tool), and that would run fully automated during authentic classroom use \citep[e.g.,][]{ekstromDualRoleHumanoid2022}. As for passive stakeholder involvement, these studies have commonly collected diverse log data and usability questionnaires to understand users' attitudes towards the use of automated intelligent support, such as the work presented by \citet{wilsonAutomatedFeedbackAutomated2021}, in which teachers and students were asked to evaluate a fully-automated writing score system using usability and attitude questionnaires.

% Q1
Lastly, systems in \textbf{Q1} that feature low human control and automation had an equal proportion of active and passive involvement, at 6\%. An active involvement is exemplified in \citeauthor{garcia-ruizParticipatoryDesignSonification2022a}'s (\citeyear{garcia-ruizParticipatoryDesignSonification2022a}) study, in which participatory design activities were conducted with students (e.g., initial exploration, focus group discovery, and collaborative prototyping with experts) to design a novel visualisation tool with a limited configuration setting. 
In contrast, \citeauthor{fernandez-nietoModellingSpatialBehaviours2021}'s (\citeyear{fernandez-nietoModellingSpatialBehaviours2021}) work reflects passive involvement, as teachers participated in interviews to share their sensemaking on a low-fidelity dashboard prototype, in which the prototype was manually generated, and the outcomes were not part of further studies. 
In summary, education stakeholders were more involved in Q4 than in other quadrants, indicating active stakeholder involvement in influencing design outcomes on systems considering end-user agency and using AI or advanced computer automation.

\begin{figure*}[h!]
    \centering
    \includegraphics[width=0.9\textwidth]{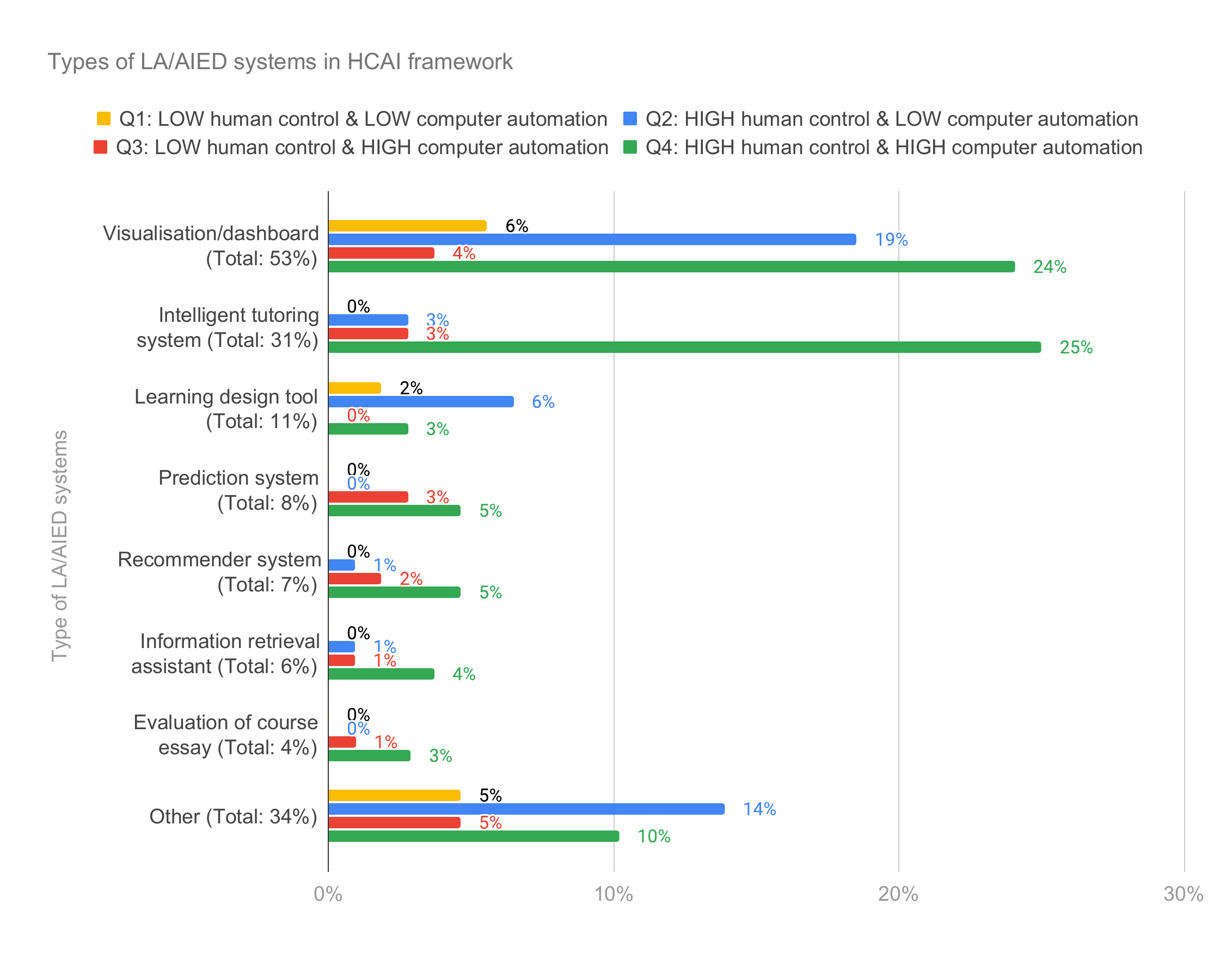}%
    \caption{Distribution of human-centred LA/AIED systems on the two-dimensional HCAI framework by type of system. Each colour represents different quadrants in the HCAI framework (Q1-Q4, see Section \ref{sec:hcai-quadrant}).}
    \label{fig:rq2-edutechs}
\end{figure*}

% TECHNOLOGIES
Concerning the different types of LA/AIED systems, as depicted in Figure \ref{fig:rq2-edutechs}, we found that the most prevalent type is visualisation or dashboards, accounting for 53\% of the total of the included studies. 

% Examples
The most prominent quadrant in this type is Q4, which comprises 24\% of the total included studies.
This type involves a high level of human control and computer automation, such as educators actively utilising and interpreting the visualised data generated by an AI-powered system which automatically make inferences by analysing complex and large amount of students' learning data \citep[e.g.,][]{alfredoThatStudentShould2023a, nazaretskyEmpoweringTeachersAI2022,conijnHumancenteredDesignDashboard2020}. 
The next most represented category is intelligent tutoring systems at a total of 31\% \citep[e.g.,][]{holstein2019Codesign, gerdesAskElleAdaptableProgramming2017, ngoonInstructorAlreadyAble2023}.
Q4 also represents the largest quadrant for this type (25\%), where high human control is given to users, such as overseeing learning activities \citep[e.g.,][]{tissenbaumSupportingClassroomOrchestration2019}, accepting or rejecting automated interventions or recommendations from the system \citep[e.g.,][]{dimitriKeepMeLoop2022}, and making real-time adjustments \citep[e.g.,][]{lawrenceHowTeachersConceptualise2023}.

% About LD and other types
Next, learning design support for teachers is represented by a total of 11\%. 
The most prominent quadrant for this type is Q2 at 6\% \citep[e.g.,][]{michosInvolvingTeachersLearning2020, vezzoliInspirationCardsWorkshops2020, pishtariMultistakeholderPerspectiveAnalytics2021}, which favours high human control, and Q4 is the second highest at 3\% 
\citep[e.g.,][]{rodriguez-trianaAdaIblLessons2021,alboKnowledgeBasedDesignAnalytics2022, kaliisaCombiningCheckpointProcess2020}, which may indicate learning design systems have lacked computer-automation.
Learning design systems commonly consider high human control, as exemplified by the agency of learning designers in applying the inquiry process where they possess the autonomy to actively shape and refine learning experiences \citep[e.g.,][]{pishtariMultistakeholderPerspectiveAnalytics2021,rodriguez-trianaAdaIblLessons2021}.  

The prediction system type accounts for a total of 8\% and only has high computer automation (Q3 at 3\% and Q4 at 5\%). An example of Q3 of this type is in \citeauthor{ocumpaughGuidanceCounselorReports2017}'s (\citeyear{ocumpaughGuidanceCounselorReports2017}) study, where the system was trained on historical students' data to identify patterns, make predictions on students' engagement, and produce non-configurable early warning reports for educators to interpret, without having any further control on the system. On the other hand, an example in Q4 can be seen in \citeauthor{duanTransparentTrustworthyPrediction2023}'s (\citeyear{duanTransparentTrustworthyPrediction2023}) study, where the teachers have more control over configuring the model and representation of prediction results to identify at-risk students, which ensures predictions are trustworthy and can support proper interventions.
Evidently, no prediction systems are at low-level computer automation, as evidenced by no studies found in Q1 and Q2, since all prediction systems require a high level of computer automation (i.e., AI models) to operate. 

The remaining three types---recommender system, information retrieval assistant, and evaluation of course essays---have Q4 as the most prominent quadrant. 
Recommender systems are represented by total of 7\%, with 5\% in Q4, that includes course recommender system \citep[e.g.,][]{changAIPleaseHelp2023}, learning resources recommender \citep[e.g.,][]{ruiz-callejaInfrastructureWorkplaceLearning2019}, and personalised learning recommendations \citep[e.g.,][]{khosraviRiPPLECrowdsourcedAdaptive2019}.  
Information retrieval assistant comprises total of 6\%, with 4\% in Q4 \citep[e.g.,][]{bonnatDigitalCompanionMonitor2022a} and evaluation of course essay 4\%, with 3\% in Q4 \citep[e.g.,][]{leeHumanCentricAutomatedEssay2023a}. Systems in Q4 across these three types have similarities where automation or an intelligent system relies on teachers' or students' oversight to interpret, make judgments, and contextualise the information from the provided interface to take actions in their teaching or learning environment close to real-time.

The other types of educational technologies accounted for smaller percentages, ranging from 1\% to 5\% of the total studies, adding up to 33\%. 
These include  game-based learning system \citep[e.g.,][]{tenorioExploringDesignConcepts2022,shuteDesignDevelopmentTesting2021}, online learning system \citep[e.g.,][]{martinezImpactUsingLearning2020}, mixed-reality systems such as augmented-reality and virtual-reality \citep{bonnatDigitalCompanionMonitor2022a, zwolinskiExtendedRealityEducation2022a, kangARMathAugmentingEveryday2020}, and social robots \citep[e.g.,][]{ekstromDualRoleHumanoid2022,muldnerDesigningTangibleLearning2013}.
In studies about social robots, such as in \citeauthor{ekstromDualRoleHumanoid2022}'s (\citeyear{ekstromDualRoleHumanoid2022}) study, Wizard of Oz tended to be employed in which the system was manually controlled by researchers in the background, but students still had full control of the interaction, such as having a flexible conversation with a robot to develop empathy and communication skills.
Among these, the most notable quadrant when aggregated was Q2 (14\%), in which students had full control when interacting with the system (i.e., playing learning games \citep{wangHowDoesOrder2019} or interacting with peers in online learning \citep{cukurovaInteractionAnalysisOnline2017}), which was manually prepared with a pre-defined set of rules towards intended learning outcomes, without requiring advanced automation.

In summary, regardless of the level of computer automation, each type of LA/AIED system has already considered a high level of human control, indicated by a higher percentage of studies categorised in Q2 or Q4 for each type. These findings suggest that most reviewed human-centred LA/AIED systems have been designed to empower users with the support of human oversight.

% RQ4: How are researchers discussing safety, reliability and trustworthiness in learning analytics systems?
\subsection{Safety, reliability, and trustworthiness discussion (RQ4)}
\label{res:rq4}

Of the total number of human-centred LA/AIED systems' studies reviewed, 54\% contained a discussion or consideration of the HCAI principles of \textbf{safety}, \textbf{reliability}, and \textbf{trustworthiness} to some extent. The individual proportions were 35\%, 37\%, and 25\%, respectively. Next, we describe the topics that emerged from each principle as a result of our thematic analysis.
% \textit{'\%'} denotes the percentage of total studies in which a topic was found. 

% Discussion of safety	(COLUMN V)
\subsubsection{Safety}

% Data Privacy
\textbf{Data privacy} emerged as the most topic (18\%), emphasising its significance when designing and implementing safe LA/AIED systems. Data privacy refers to the protection and control of personal and sensitive information of students, teachers, or any other individuals involved in the learning process. These studies involved discussions about safeguarding data related to personal identification \citep[e.g.,][]{kivimakiCurricularConceptMaps2019, ngoonInstructorAlreadyAble2023}, academic performance \citep[e.g.,][]{echeverriaCollaborationTranslucenceGiving2019}, learning progress \citep[e.g.,][]{maGlanceeAdaptableSystem2022}, and other sensitive details collected and processed by human-centred LA/AIED systems \citep[e.g.,][]{santosGoalOrientedVisualizationsActivity2012}.
% This data privacy is closely related to how transparent the information exposed such personal information.
Notably, discussion about privacy could influence stakeholders' willingness to adopt these systems in practice \citep[e.g.,][]{maGlanceeAdaptableSystem2022, garcia-ruizParticipatoryDesignSonification2022a}. 

% privacy in data sharing.
The discussion of \textbf{data sharing} primarily focused on strategies to overcome privacy concerns (7\%). Four studies advocated for anonymity features when sharing data as measures to protect participants' identities from others, promoting a more open discussion of learning activities \citep[e.g.,][]{echeverriaCollaborationTranslucenceGiving2019, conijnHumancenteredDesignDashboard2020, satoGroupnamicsDesigningInterface2023a, barreirosStudentsFocusMoving2023}.
These anonymity features could also help students recall and discuss performed actions without feeling judged by their peers \citep[i.e.,][]{echeverriaCollaborationTranslucenceGiving2019}.
Moreover, six studies discussed safety strategies for data sharing as part of their design methodologies. 
Two of them incorporated design principles such as "Privacy by Design" criteria \citep[i.e.,][]{bonnatDigitalCompanionMonitor2022a}, allowing privacy and design to co-exist, and "Risk Communication Principles" \citep[i.e.,][]{ocumpaughGuidanceCounselorReports2017}, allowing communicating the risk involved and the certainty when presenting data in a way that both experts and non-experts (i.e., students) can easily understand during design activity. 
Other studies leveraged co-design methods with teachers \cite[e.g.,][]{ ngoonInstructorAlreadyAble2023, pishtariMultistakeholderPerspectiveAnalytics2021} and students \citep[e.g.,][]{leePrototypingSpatialSkills2022a, garcia-ruizParticipatoryDesignSonification2022a}, to explore the type of data that can be shared with other users without breaching their privacy. 

% data collection
We found some discussions about \textbf{data collection} impact and procedures (10\%).
Two studies reported conflicting opinions from participants regarding the benefits and concerns associated with collecting and utilising student data \citep[i.e.,][]{wangCoDesigningAIAgents2022, kivimakiCurricularConceptMaps2019}. For example, \citet{wangCoDesigningAIAgents2022} reported that several students expressed worries about potential privacy breaches and discomfort in asking questions in an online forum (due to being tracked and continuous data collection), but they wanted to have a more personalised learning experience from the system. It exemplifies how students may want something beneficial for their learning with an intelligent system but are unaware that such a system requires data collection. 
% data collection procedures
It might be necessary to establish clear communication between researchers and participants through informed consent. Five studies highlighted the importance of informed consent procedures to ensure transparent data collection \citep[i.e.,][]{conijnHowProvideAutomated2020, khosraviRiPPLECrowdsourcedAdaptive2019, wangCoDesigningAIAgents2022, dequinceyStudentCentredDesign2019, ngoonInstructorAlreadyAble2023}. 
By acquiring explicit consent from students (e.g., opt-in/out process) and allowing users to control data access and explaining how the system operates \citep[i.e.,][]{bonnatDigitalCompanionMonitor2022a, wangCoDesigningAIAgents2022}, human-centred LA/AIED systems could potentially deliver a sense of safety to users. 
Yet, \citet{dequinceyStudentCentredDesign2019} argued that some participants may overlook the privacy policy of the informed consent process because they may not have read the information thoroughly. 

Moreover, \textbf{monitoring and surveillance} based on student data were discussed in four studies \citep[i.e.,][]{pozdniakovQuestiondrivenDashboardHow2022,zhouInvestigatingStudentsExperiences2021, alfredoThatStudentShould2023a, tsai2020empowering}. 
For instance, \citet{zhouInvestigatingStudentsExperiences2021} reported the experiences, privacy concerns, and impacts of providing students awareness for using monitoring tools in a collaborative learning setting (i.e., remote group meetings). They reported changes in students' behaviours, such as half of students feeling more motivated to engage and be more productive in the group conversations, while half of students still felt intense pressure and uncomfortable being monitored. 
In another study, \citet{alfredoThatStudentShould2023a} further advised against using physiological data modelling and visualisation for surveillance purposes, including scenarios like exams and regular classrooms, as well as measuring students' performance.
Another study by \citet{ngoonInstructorAlreadyAble2023} suggested that institutions that plan to adopt smart classroom systems with continuous monitoring features should conduct regular evaluations with both students and teachers, such as addressing privacy concerns through an iterative co-design process. 
Finally, a small number of studies (3\%) proposed strategies to ensure secured data access (\textbf{data security}), such as two-factor authentication \citep[i.e.,][]{wangCoDesigningAIAgents2022}, using third-party software to establish privacy and security policies \citep{santosGoalOrientedVisualizationsActivity2012}, and allowing limited data accessibility and visibility to relevant stakeholders (i.e., students) intended only for academic purposes \citep{martinezImpactUsingLearning2020}.

\subsubsection{Reliability}
% Discussion of reliability	(COLUMN W)
In terms of reliability, we found that this principle is often addressed in terms of the accuracy of the system, the potential bias of these systems, and strategies to produce reliable data. We describe each of these topics below. 

From the set of studies, we found \textbf{accuracy} (14\%) as a key aspect to evaluate the system's reliability. These studies often consider two factors that could impact the accuracy and perception of the system's reliability: \textit{machine algorithm} (3\%) and \textit{human interpretation} (11\%).  
Accuracy in \textit{machine algorithm} refers to the ability of the system to provide correct/accurate outcomes. 
Two studies discussed approaches to evaluate AI model accuracy that can lead to reliable outcomes \citep[e.g.,][]{changAIPleaseHelp2023, khosraviRiPPLECrowdsourcedAdaptive2019}. For example, \citet{changAIPleaseHelp2023} used a mixed-method approach that integrated students' survey data to evaluate the accuracy of their proposed machine-learning model.
Similarly, in a study on recommender systems, \citet{khosravi2022explainable} reported initial results about incorporating both students' subjective opinions and machine-learning algorithms, which can potentially improve the accuracy of the system in determining learning resource quality.
On the other hand, accuracy in \textit{human interpretation} refers to the significance of how end-users (teachers/students) correctly interpret and understand indicators presented by the LA/AIED systems. 

Six studies reported stakeholders' responses about receiving \textbf{inaccurate} information or conclusions from a system that can introduce unreliability due to misinformation \citep[i.e.,][]{liaqatCollaboratingMatureEnglish2021, pozdniakovQuestiondrivenDashboardHow2022,shreinerInformationWonJust2022a,satoGroupnamicsDesigningInterface2023a,fernandez-nietoModellingSpatialBehaviours2021, kangARMathAugmentingEveryday2020}. 
For example, in \citeauthor{shreinerInformationWonJust2022a}'s (\citeyear{shreinerInformationWonJust2022a}) study, teachers critically analysed data visualisations to support students in learning data visualisations creation with a computer. Teachers discovered many different ways in which data visualisation generated by computers can be misleading. Specifically, teachers felt less confident in their own abilities to identify flaws and inaccuracies in computer-generated data visualisations. In their further exploration, authors reported teachers felt more confident creating data visualisations with paper and pencil over using computers. As a result, only a small number of the teachers who were supposed to use this system have actually incorporated them into their classrooms.
Moreover, seven studies reported how stakeholders questioned the accuracy of the results from LA/AIED systems \citep[i.e.,][]{Ahn2021Apr,zhouInvestigatingStudentsExperiences2021, shreinerInformationWonJust2022a, barreirosStudentsFocusMoving2023}, specifically in addressing social needs to promote social learning development \citep[i.e.,][]{wangCoDesigningAIAgents2022}, emotional support needs \citep[i.e.,][]{liaqatCollaboratingMatureEnglish2021}, and stress \citep[i.e.,][]{alfredoThatStudentShould2023a}. For instance, in a couple of studies \citep[i.e.,][]{Ahn2021Apr, zhouInvestigatingStudentsExperiences2021}, study participants doubted whether the data they saw in the system's interface reflected what occurred in the learning activity. 
An example from \citeauthor{kangARMathAugmentingEveryday2020}'s (\citeyear{kangARMathAugmentingEveryday2020}) study addresses this unreliability issue by providing students with control to correct the mistakes made by AI (i.e., incorrectly assigned labels to an interface). At the same time, it promotes students' agency over the tool to acknowledge that the system can be unreliable and corrections may be needed during use. 

Another topic was \textbf{human bias} (4\%), which refers to systematic and unfair preferences influencing the outcomes, decisions, or interpretations generated by systems \citep[e.g.][]{fernandez-nietoClassroomDandelionsVisualising2022, crainVisualizingTopicsOpinions2022a, pozdniakovQuestiondrivenDashboardHow2022,alfredoThatStudentShould2023a}.
Two studies highlighted the possibility of bias risk in LA/AIED systems due to the perceptions and preferences of users who have control over them \citep[e.g.][]{fernandez-nietoClassroomDandelionsVisualising2022,crainVisualizingTopicsOpinions2022a}, arguing that excessive human control could lead to over-interpretation of data.  
The other three studies reported that insights formulated based on inaccurate visual analytics might be prone to interpretive bias \citep{pozdniakovQuestiondrivenDashboardHow2022,alfredoThatStudentShould2023a,fernandez2022beyond}.
% bias in design
We also gathered several discussions about human bias in the design process, as examined by four studies (4\%).
Two studies explicitly discussed the risk of bias in the HCD process \citep[i.e.,][]{vinellaFormingTeamsLearners2022,dimitriKeepMeLoop2022}. For instance, \citet{vinellaFormingTeamsLearners2022} noted that the monetary incentive given to participants when employing HCD techniques (e.g., crowdsourcing in this study) could stir and bias stakeholders from the study's intended purpose. Participants who are more focused on immediate rewards may not necessarily invest the necessary time and effort to provide insights that contribute to the long-term reliability of the system. 
However, two studies discussed the importance of balancing researchers' bias with stakeholders' needs in the design process \citep[i.e.,][]{vannaprathipIntelligentTutoringSurgical2022, fernandez2022beyond}. For instance, \citet{fernandez2022beyond} highlighted that 'end-users' (teachers/students) should be actively involved in design and evaluation processes to perceive their real needs in order to minimise researchers' bias. When researchers or designers work in isolation, they may unintentionally introduce their own assumptions and biases into the system. Diverse perspectives are considered when involving end-users, leading to a better alignment with challenges found in authentic educational settings that can be more reliable for end-users use.

Furthermore, another topic that emerged is related to the \textbf{strategies to produce reliable data} (5\%).
Three studies discussed how they ensured the reliability of the data from their studies \citep[i.e.,][]{martinezImpactUsingLearning2020,leeHumanCentricAutomatedEssay2023a,ocumpaughGuidanceCounselorReports2017}. 
For instance, \citet{martinezImpactUsingLearning2020} ensured the reliability of their results by applying criteria such as data security and confidentiality in data collection, triangulating data from multiple sources, and reviewing the analysis process for each design phase.
Two studies reported that the methodological reliability for involving stakeholders in the design process could be influenced by the quality of evaluations from stakeholders \citep[i.e.,][]{ahnDesigningContextReaching2019,zhouInvestigatingStudentsExperiences2021}.
For instance, \citet{ahnDesigningContextReaching2019} highlighted that methodological reliability from a study might be influenced by how well the involved stakeholders understand the system's \textit{validity} in an educational setting. Validity refers to how well the expectations of those involved match the intended purpose of the system, which promotes consistency and relevancy in evaluating the system to produce reliable data. 
Moreover, three studies addressed the issue of system validity, noting that while these systems may function effectively in laboratory settings, their sustainability in real-world scenarios remains uncertain \citep[i.e.,][]{zhangStoryBuddyHumanAICollaborative2022, gibsonReflectiveWritingAnalytics2017, huITalkISeeParticipatory2022a}. They suggested the need for \textit{ecological validity} (the extent to which the systems and findings from studies can be generalised and applied to authentic educational settings) by incorporating intended stakeholders' perspectives to design a more reliable system. 

\subsubsection{Trustworthiness}
Three emerging topics were identified in relation to the trustworthiness principle, such as \textit{trust} (14\%), \textit{transparency} (9\%), and \textit{accountability} (3\%) in LA/AIED systems.

\textbf{Trust} was discussed in several studies (14\%). This concept was contextualised in how the stakeholders \textit{perceive} trustworthiness when using or designing LA/AIED systems \citep[e.g.,][]{duanTransparentTrustworthyPrediction2023, ngoonInstructorAlreadyAble2023, fernandez-nietoClassroomDandelionsVisualising2022,ekstromDualRoleHumanoid2022, barreirosStudentsFocusMoving2023, oogeSteeringRecommendationsVisualising2023a}.
In terms of evaluating trust of such LA/AIED systems, we have mixed findings on stakeholders' perceived sense of trust. %on trust in the current systems.
Four studies reported that stakeholders trusted the outcomes of AI or computer automation \citep[i.e.,][]{liaqatCollaboratingMatureEnglish2021, gibsonReflectiveWritingAnalytics2017,khosraviRiPPLECrowdsourcedAdaptive2019, maGlanceeAdaptableSystem2022}. 
For instance, \citet{liaqatCollaboratingMatureEnglish2021} reported that learners trusted automated feedback more than peer feedback, and \citet{maGlanceeAdaptableSystem2022} noted that teachers were inclined to trust the system when the displayed data matched their expectations.
On the other hand, four studies reported stakeholders were less trusting of outcomes from automation \citep[i.e.,][]{nazaretskyInstrumentMeasuringTeachers2022, wangCoDesigningAIAgents2022,oogeSteeringRecommendationsVisualising2023a}, especially when the systems utilised emotion or stress data \citep[i.e.,][]{alfredoThatStudentShould2023a}. 
For instance, \citet{nazaretskyInstrumentMeasuringTeachers2022} reported that teachers were less trusting of the AI-powered system compared to receiving advice from peer teachers or experts. 
In another study, \citet{alfredoThatStudentShould2023a} reported teachers doubted the system's ability to infer students' physiological states accurately due the inherent complexity in modelling affective constructs. They suggested additional contextual information (e.g., explainable data) that could support making inferences and explaining this data back to students, especially if the intention is to use these systems in authentic settings.
Interestingly, \citet{oogeSteeringRecommendationsVisualising2023a} argued that plainly giving learners a mechanism to control a recommender system, such as an ability to reveal the number and detail of recommendations during a learning activity, does not necessarily increase their trust in using the system. Instead, the authors highlighted this kind of human control could promote awareness of the algorithm behind computer automation, encouraging learners to reflect on their actions which may foster long-term trust in the system.
Together, these results provide important insights into further investigating the nuanced dynamics of trust between stakeholders and the system, which may vary depending on who the end-user is (e.g., teachers or students).

% 2) Mechanisms to cultivate/foster trust (these could be part of the design - for example - in the evaluation of these systems)
Moreover, four studies discussed several approaches to cultivate stakeholders' trust in the system \citep[i.e.,][]{Ahn2021Apr,wilsonAutomatedFeedbackAutomated2021, sulemanNewPerspectiveNegotiationBased2016, gibsonReflectiveWritingAnalytics2017}.
For instance, \citet{wilsonAutomatedFeedbackAutomated2021} reported that balancing the presented information's accuracy with explainable information could foster learners' trust in the system. The authors ensured that the system is available to provide output whenever a learner requests feedback, but the output may be less accurate. This drawback was balanced by providing learners with an explanation that is more connected to the intended pedagogical component (i.e., formative feedback). It exemplifies an approach to improving trustworthiness when using human-centred LA/AIED systems by improving interpretability through explainability and system availability. 
\citet{gibsonReflectiveWritingAnalytics2017} emphasised that stakeholders' trust in LA systems can be developed through reciprocity, meaning there exist continuous design processes that involve active evaluation and the ability to influence the programming code inside the system. The study highlights the importance of cultivating trust between stakeholders and systems to foster trust in using them in the authentic learning context.

% Transparency
% \citep[e.g.,][]{pozdniakovQuestiondrivenDashboardHow2022, oogeSteeringRecommendationsVisualising2023a, conijnHowProvideAutomated2020, dollinger2019working, khosraviRiPPLECrowdsourcedAdaptive2019, maGlanceeAdaptableSystem2022}
\textbf{Transparency} (9\%) was the second most discussed topic in the trustworthiness principle.
Transparency refers to the clear and understandable way the system gathers and manages information to build trust among stakeholders who use or contribute to the LA/AIED system's design.
Five studies discussed the transparency strategies that could foster trust, built around the user's perceptions and control \citep[i.e.,][]{shuteDesignDevelopmentTesting2021, Ahn2021Apr,khosravi2022explainable,sulemanNewPerspectiveNegotiationBased2016, dollingerWorkingTogetherLearning2019}. 
\citet{shuteDesignDevelopmentTesting2021} highlighted the importance of users' perceptions and their significant role in designing and testing the LA/AIED system impacting trust and engagement. 
\citet{khosraviRiPPLECrowdsourcedAdaptive2019} reported that educating students on how their knowledge status is computed by AI and giving them access to the learning model (i.e., opening up the black-box AI models) could ultimately increase their trust and motivation to learn. 
\citet{Ahn2021Apr} suggested that communicating the perception and intention of data use can foster trust through ongoing interpersonal interactions.
Similarly, \citet{dollingerWorkingTogetherLearning2019} emphasised that transparent communication and clear research documentation are fundamental for building trust between stakeholders and establishing trustworthiness.
This finding highlights that trust and engagement in LA/AIED systems can be positively influenced when users perceive transparency in knowledge computation and transparent communication between stakeholders and researchers is encouraged. 

% Accountability
The concept of \textbf{accountability} refers to individuals or systems being responsible for their decisions and the consequences of their actions. This is discussed in only four studies, taking a perspective from system usage \citep[i.e.,][]{zhouInvestigatingStudentsExperiences2021, duanTransparentTrustworthyPrediction2023,lawrenceHowTeachersConceptualise2023} and design \citep[i.e.,][]{gibsonReflectiveWritingAnalytics2017}. 
The study by \citet{zhouInvestigatingStudentsExperiences2021} sees accountability from a system usage perspective. For them, accountability is not just about the teachers overseeing the system; it also involves students taking responsibility for adjusting their learning strategies based on the feedback they receive from the AI-powered system. This suggests that the accountability of the AI-powered system usage is not solely dependent on teachers but also on the students' willingness to use the information provided by the system in their learning.
The importance of accountable use of an AI-powered system is also investigated in \citeauthor{duanTransparentTrustworthyPrediction2023}' (\citeyear{duanTransparentTrustworthyPrediction2023}) study, where authors highlighted the need for trustworthy and transparent AI algorithms that align with the needs of stakeholders involvement in real educational settings. 
The authors recommended exploring explainable AI technologies capable of dynamically incorporating subject-matter experts’ (i.e., teachers') contextual insights during the learning process for generating more accountable interventions aimed at helping students accomplish their learning goals. 
This kind of teachers' interventions are discussed in \citeauthor{lawrenceHowTeachersConceptualise2023}'s (\citeyear{lawrenceHowTeachersConceptualise2023}) study. Authors highlighted that teachers felt responsible for classroom orchestration decisions made with the assistance of the AI-powered system, thus suggesting that teachers should be encouraged to trust their own judgement and decision-making ability over the AI. Teachers explained they tended to override the system's recommendations with their own judgement, indicating teachers may still hesitate to trust and follow AI suggestions completely. This study may suggest that when teachers have control over the system, the control features force them to be responsible for the system's effectiveness since they are accountable for their decisions, which may give them an extra burden to use in practice.

In contrast, \citet{gibsonReflectiveWritingAnalytics2017} define accountability from a design standpoint study. The expectation was for teachers to exhibit a higher level of accountability, implying that they should be more critical during the evaluation phase of the design process. However, it was observed that the teachers did not meet this expectation and their evaluations were not as critical as anticipated. This highlights a need to minimise assumptions about stakeholders' accountability in the design process.
In summary, effective exploration of accountable usage in human-centred LA/AIED systems requires active participation from both teachers and students. However, the limited literature from a design perspective shows a gap in our understanding of the accountability of stakeholders when involved in the design process. Each study interprets its meaning differently, primarily focusing on accountability in system usage.

\section{Discussion}
\subsection{Current state of Human-centred LA/AIED systems}

% RQ1
% THEMES AND current CHALLENGES/GAP
The results from \textbf{\textit{RQ1}} indicate that a high number of university/college studies align with other LA/AIED reviews \citep{sarmiento2022participatory,Granic2022Aug}. 
Yet, there are still opportunities to explore human-centred LA/AIED systems in supporting informal learning \citep[e.g.][]{zhangStoryBuddyHumanAICollaborative2022,headToneWarsConnectingLanguage2014}, as noted by \citet{buckingham2019human}. 
Moreover, the HCD techniques most commonly used in designing human-centred LA/AIED systems that have considered stakeholders' voices include interviews, co-design sessions, and prototype validation, with few studies using other techniques such as surveys, workshops, and personas. 
Despite traditional techniques (interviews and questionnaires) being helpful, this area requires more innovative HCD techniques to leverage AI's potential challenges and benefits.
For example, other techniques and methods that mimic human-AI interaction (i.e., Wizard of Oz \citep{vinellaFormingTeamsLearners2022} or technology probes \citep{satoGroupnamicsDesigningInterface2023a}) could be helpful to investigate how users interact with systems features that require complex automation without the burden of full implementation \citep{lawrenceHowTeachersConceptualise2023,echeverriaExploringHumanAI2020}. However, researchers should consider the trade-off between high-fidelity interaction and resources needed, as these techniques and methods may require more resources and expertise than low-fidelity prototypes \citep{muldnerDesigningTangibleLearning2013}.

% -- Lack of Phase 1 and 5 and Quantitative research had not been employed in Phase 1; 
The lack of human-centred LA/AIED systems in Phase 1 (18\%) and Phase 5 (9\%) highlights a need for greater attention to the inception and monitoring phases.
Engaging stakeholders at the earliest and latest design phase can provide valuable perspectives that ensure the designed system aligns with real-world learning needs and concerns \citep{buckingham2019human} and becomes sustainable solutions \citep{Yan2022Mar}. 
A closer look at the planning, scoping, and definition design phase (Phase 1) indicates quantitative methodologies are scarce. % suggesting that numerical data-driven analysis could be lacking in establishing research initiatives. 
Integrating quantitative methodologies, such as surveys or crowdsourcing, could offer an objective, systematic approach to gathering and interpreting data that could capture ideas at a large scale for subsequent design phases. For example, researchers can conduct surveys on end-users (teachers/students) to assess potential risks or feasibility, which can be cost-effective and scalable; otherwise, it may require more effort and resources when stakeholders are involved in subsequent design phases \citep{langLearningAnalyticsStakeholder2023}. Researchers may also consider analysing historical or public data to develop relevant AI-powered features or models depending on the research objectives \cite[i.e., by following data-driven design approaches,][]{Gorkovenko2020futuredatadriven}. 
Moreover, the limited literature in the launch and monitoring design phase (Phase 5) could suggest a gap in reporting practices about LA/AIED systems post-deployment in authentic learning environments \citep{martinez-maldonadoHumancentredLearningAnalytics2023} or a lack of longitudinal studies that manage to deploy human-centred LA/AIED systems sustainably.  

Therefore, evidenced by our findings, we recommend:
\begin{itemize}
    \item \textbf{Research Scope:} Expand human-centred LA and AI in Education research beyond the university and K-12 levels to include vocational education, workplace training, and informal learning.
    \item \textbf{Inception \& Deployment Phases:} Pay greater attention to Phases 1 and 5 of the design process to balance stakeholder involvement, pedagogical focus, and research objectives at both inception (phase 1) and monitoring (phase 5) stages.
    \item \textbf{Quantitative Approaches:} Employ quantitative methods in the planning phase to help reach stakeholder consensus and be resource-effective. Consider large-scale surveys, crowdsourcing, quantified risk analysis, or leveraging public datasets on AI-powered features.
    \item \textbf{In-the-Wild Studies:} Report more in-the-wild studies during Phase 5 to understand the impact of human-centred LA/AIED systems on teaching or learning experiences in real-world practices. Combine this with existing human-centred approaches and design methodologies \cite[e.g., design-based research by][]{Reimann2016}.
    \item \textbf{Innovative HCD Techniques:} Use innovative Human-Centered Design techniques beyond interviews and questionnaires to leverage AI’s potential challenges and benefits. Consider techniques that mimic human-AI interaction, such as Wizard of Oz.
\end{itemize}

\subsection{Stakeholder involvement in Design Phases}
Regarding \textbf{\textit{RQ2}}, lower active stakeholder involvement in evaluating and refining human-centred LA/AIED systems can overlook design flaws that may persist in authentic educational scenarios (see Phase 4).
Including their voices in the refinement process, such as conducting post-hoc reflection interviews to capture their experiences for using a system after pilot studies \citep{martinez-maldonadoLATUXWorkflowDesigning2015}, can provide an opportunity for end-users to reflect and re-align systems' features with their values, preferences, and needs \citep{chenValueSensitiveLearningAnalytics2019} before the system is finalised and deployed in real-world settings. 
% - Human-in-the-loop is required in all stages of design:
This active stakeholder involvement should be considered in all design phases, which have currently been considered in Phases 1, 2, 3, and 5.
Incorporating a "human-in-the-loop" practice in all design phases can be an approach to tap into education stakeholders' expertise, ensuring a nuanced grasp of context, end-user needs, and ethical considerations \citep{duanTransparentTrustworthyPrediction2023}. 
Allowing end-users to share their voices can also foster a sense of ownership, which aligns with the main goal of human-centred LA/AIED systems to empower end-users  \citep{usmaniHumanCenteredArtificialIntelligence2023,barreirosStudentsFocusMoving2023}.
Ultimately, this active involvement will foster adaptive and human-centric designs that align more effectively with real-world complexities and requirements \citep{buckingham2019human, martinez-maldonadoHumancentredLearningAnalytics2023}.
Indeed, it is crucial to acknowledge that a balanced combination of HCD techniques and design phases for involving stakeholders in educational settings, such as availabilities, policies, and resources, remains an open question that requires further exploration.
Considering these benefits, this finding highlights the need for stakeholder involvement in all design phases of human-centred LA/AIED systems to co-create systems that can meet real-world educational needs \citep{dollingerWorkingTogetherLearning2019}.

The extent of stakeholder involvement indicates that students and teachers play a pivotal role in the design process as the most frequently engaged stakeholders (see Fig. \ref{fig:rq1-stakeholders-involvement}). 
Despite students having the highest representation in the overall participant count (71\%), the data highlighted their active involvement in the design process was relatively low (19\%).
Teachers may have expertise and pedagogy knowledge (e.g., how students learn, effective instructional strategies, and a supportive learning environment). 
In contrast, students may be just consumers of these systems or partake in another role in the design, such as being observed when exploring the system or participating in experiments to test the newly developed system without giving input or voice to the design. 
It could be caused by several challenges, such as communication, students' narrowed perspectives or knowledge, monetary incentives \citep{vinellaFormingTeamsLearners2022} and privacy concerns, especially in K-12 education \citep{Bond2023Mar} (see Section \ref{res:rq4}). 
This may lead to a heavy reliance on teachers and other experts, which invites further investigation. 
Same teachers or experts may be more available to contribute to the design process from the beginning until the end of the project than students, which commonly occurs in the longitudinal study \citep[e.g.,][]{huttBreakingOutLab2021}.

The challenges above show that students' contributions tend to be punctual (occurring as one time-off) and sporadic (happening sometimes and inconsistently). 
These uncertainties signal a need for a more inclusive approach to the design process due to the underutilisation of students’ expertise and voices in the design process of human-centred LA/AIED systems.
To break this pattern, efforts should focus on recognising the value of students' unique insights \citep{dollingerWorkingTogetherLearning2019, Alfredo2024slade}, empowering them through agency \citep{hooshyarLearningAnalyticsSupporting2023c}, and addressing communication problems, such as by clearly stating the research outcomes and benefits of participating in the design activities \citep{sladeLearningAnalyticsEthical2013}.
It is also essential to encourage collaboration among students, teachers, researchers, and other stakeholders (e.g., developers, designers, and administrators) while establishing transparent communication about the purpose and benefits of student involvement \citep{martinez-maldonadoHumancentredLearningAnalytics2023}. This multifaceted approach aims to collaboratively create a more inclusive and effective system \citep{dollingerWorkingTogetherLearning2019}.

Overall, we found a gap in which students' active involvement is still limited, implying students can be considered as underrepresented stakeholders \citep{martinez-maldonadoHumancentredLearningAnalytics2023}. 
\citeauthor{martinez-maldonadoHumancentredLearningAnalytics2023} recommended building robust relationships with these underrepresented stakeholders, offering compensation for their time in contributing to design activities and employing inclusive design toolkits that promote inclusivity and diversity.
It urges researchers to be more proactive in including students' voices because their learning experiences will be the most impacted by these technologies \citep{kitto2018embracing}.
We further argue that ensuring meaningful student engagement in the design process can enhance the relevance and effectiveness of adopting human-centred LA/AIED systems, promoting a more student-centred approach \citep{dequinceyStudentCentredDesign2019}. 

Therefore, we recommend:
\begin{itemize}
    \item \textbf{Active Stakeholder Involvement}: Employ co-creation practices such as co-design, participatory design, and value-sensitive design. Emphasise active collaboration in all design phases and the central role of end-users and directly affected stakeholders (including students) in shaping outcomes.
    \item \textbf{End-User Needs and Preferences}: Ensure the system's outcomes meet the needs and preferences of the end-users, particularly teachers and students.
    \item \textbf{Student Involvement}: Address students’ passive involvement in the design process. Recognise that human-centred LA/AIED systems can significantly impact students’ learning experiences and outcomes.
    \item \textbf{Clear Communication}: Ensure all stakeholders are well-informed about the benefits and challenges of involving students in the design process.
\end{itemize}

% RQ3
\subsection{Human Control and Computer Automation}
Regarding \textbf{\textit{RQ3}}, by employing the two-dimensional HCAI framework \citep{shneidermanHumanCenteredAI2022a} as an analysis lens, the current SLR revealed that current human-centred LA/AIED systems have already prioritised human control with a significant percentage falling in this category (Q2=32\% and Q4=47\%). 
Human control is prioritised so students and teachers can be responsible for their learning and teaching practices. With AI-powered systems' help and high human control (Q4), students can have personalised learning experiences, and teachers can provide guidance and support \citep{doganUseArtificialIntelligence2023}. This finding also indicates the need for further exploration of frameworks for collaboration between humans and AI in the LA/AIED systems \citep{holsteinConceptualFrameworkHuman2020}. This approach stands in line with the objectives of HCAI \citep{shneidermanHCAI2020, usmaniHumanCenteredArtificialIntelligence2023, renzReinvigoratingDiscourseHumanCentered2021}.

Regarding stakeholder involvement in the HCAI framework, notably, Q3 (low human control and high computer automation) has shown more passive stakeholder involvement (6\%). 
The possible reason for this passive involvement, as highlighted by \citet{Nazaretsky2022Jul}, could be a lack of technical expertise among the stakeholders. Systems in Q3 often have simpler interfaces, resulting in lower human control but it can be more complex behind the interface. It may require a certain understanding of technical knowledge, such as computer science or AI literacy, to understand its automation. If teachers and students lack this expertise, they might find it challenging to provide meaningful input in the design process, resulting in them taking a more passive role.
Yet, stakeholders' involvement should not only be considered for human-centred LA/AIED systems that offer some sort of end-user interface but also in the design of systems that fully automate educational actions. 
For systems with low human control and high computer automation, it is important to ensure they are transparent, trustworthy, and user-friendly, as explored in \citep{duanTransparentTrustworthyPrediction2023}'s study (Q4) to balance it with giving more human control.
Another example is a ZoomSense system with high computer automation in \citeauthor{zhouInvestigatingStudentsExperiences2021}'s (\citeyear{zhouInvestigatingStudentsExperiences2021}) and \citeauthor{pozdniakovQuestiondrivenDashboardHow2022}'s (\citeyear{pozdniakovQuestiondrivenDashboardHow2022}) studies. Initially,  \citeauthor{pozdniakovQuestiondrivenDashboardHow2022}'s (\citeyear{pozdniakovQuestiondrivenDashboardHow2022}) study focused on computer automation with AI but had limited control (Q3). It was later updated by creating a query-driven dashboard that allows teachers to control the online classroom in \citet{pozdniakovQuestiondrivenDashboardHow2022}'s study (Q4).

Nonetheless, we also found that systems in quadrant Q2 had more passive stakeholder involvement (19\%). 
Since systems in Q2 are intended to provide an end-user interface that allows teachers or students to take control and make meaningful decisions about their teaching or learning process, it can be seen as short-sighted to neglect active stakeholder involvement just because AI or computer automation is unavailable.
\citet{buckinghamshumLearningAnalyticsAI2019} and \citet{dollingerWorkingTogetherLearning2019} argued that designing human-centred LA/AIED systems requires a well-thought-out stakeholder engagement strategy that considers the diverse needs and values regardless of their level of computer automation.

In intelligent tutoring systems, the control is usually given to end-users to a limited extent since these systems commonly provide a set of problems to students, adapting to their prior and current knowledge \citep{lawrenceHowTeachersConceptualise2023,schulzTutoringSystemSupport2022}. 
While teachers often do not have control to choose problems for students, they are likely to take action based on those problems, such as providing additional instruction, offering feedback, or adjusting the course material in response to the problems and how students are handling them \citep{holstein2019Codesign}. 
Hence, it could be that the system's control may often be attributed to teachers rather than students. This requires further exploration to determine how much the teacher's agency has been balanced with the student's agency \citep{echeverriaExploringHumanAI2020,lawrenceHowTeachersConceptualise2023}.
Overall, most existing human-centred LA/AIED systems have been designed with a high degree of human control, as indicated by the larger proportion of systems categorised into Q2 or Q4. It further suggests that these systems are already designed to empower end-users, supported by human oversight. This aligns with the notion of striking a balance between human control and computer automation, as too much of either can lead to users being overwhelmed with options to control or mistrust since it may operate in a 'black-box' manner, respectively \citep{ozmen2023six}. 

Therefore, we recommend:

% SHORTER VERSION:
\begin{itemize}
    \item \textbf{Define End-Users}: Clearly define the intended end-users to avoid ambiguity in agency between teachers and students in highly automated systems. Develop distinct interfaces for teachers and students to enhance system usability and effectiveness.
    \item \textbf{Involve Stakeholders}: Actively engage stakeholders in the development process regardless of the system’s level of automation or AI. Consider the influence of stakeholders’ technical expertise on their preference for human control and computer automation.
    \item \textbf{Maintain Control}: In high computer automation systems (Q3 and Q4), allow teachers and students to adjust system parameters to suit dynamic changes in use. This ensures system flexibility and adaptability according to the changing dynamics of the classrooms or learning environments.
    \item \textbf{HCAI Framework as a Design Tool}: Utilise the HCAI two-dimensional framework as a design tool to assess and balance features that offer human control and computer automation in the system (update sequentially from Q1 to Q4), empowering end-users and supporting self-efficacy.
\end{itemize}

\subsection{Exploring Safety, Reliability, and Trustworthiness in Human-Centred LA/AIED Systems}
% RQ4
Lastly, regarding \textit{\textbf{RQ4}}, we summarised how researchers address safety, reliability, and trustworthiness princples in human-centred LA/AIED systems.
Finding reveals that reliability (37\%) emerges as the most prominent principle, closely followed by safety (35\%), with trustworthiness trailing at 25\%. This distribution of principles is reflected in the design and evaluation of human-centred LA/AIED systems.
While these principles are addressed to a certain degree, our review underscores the need for more robust evidence and methodological mechanisms to thoroughly understand and evaluate them. 
This is particularly true for trustworthiness, which is less represented in the current research. However, the importance of trustworthiness in the context of LA/AIED systems cannot be overstated. As highlighted by \citet{shneidermanHCAI2020}, the perceived trustworthiness of our systems can directly impact their adoption in real-world practices. If end-users (i.e., teachers or students) deem these smart systems untrustworthy, it could significantly hinder their widespread acceptance and use.
Investigating further on trustworthiness, interestingly, we identified various perspectives on accountability; on one hand, there are several discussions about the accountability of teachers and students when using the system \citep[i.e.,][]{zhouInvestigatingStudentsExperiences2021,duanTransparentTrustworthyPrediction2023,lawrenceHowTeachersConceptualise2023}, and on the other hand, one study refers to the accountability of teachers in providing design inputs during system evaluation to be trusted stakeholders. These findings underscore the complexity of accountability in the context of human-centred LA/AIED systems \citep[i.e.,][]{gibsonReflectiveWritingAnalytics2017}. It is a shared responsibility involving teachers, students, and the system itself. Each plays a crucial role in ensuring the system’s effectiveness and in building trust among stakeholders. It extends beyond the traditional notion of individuals or systems being responsible for their actions and their consequences in use. Further investigation is required as this insight could have significant implications for the design of AI-powered systems in education.

% Interconnectedness between safety, reliability, and trustworthiness.
From our findings, the principles of safety, reliability, and trustworthiness are not isolated; rather, they are interconnected and mutually reinforcing in the context of human-centred LA/AIED systems. 
Safety is a fundamental aspect that can significantly contribute to building trust. When users, especially students, feel that their data is protected and that the system operates within safe parameters, their trust in the system will likely increase \citep{Drachsler2016Apr}. It can empower users by giving them the confidence to use the system without fear of data breaches or misuse. For instance, robust data privacy measures \citep[e.g.,][]{kivimakiCurricularConceptMaps2019, ngoonInstructorAlreadyAble2023} and transparent data collection procedures \citep[e.g.,][]{conijnHowProvideAutomated2020, wangCoDesigningAIAgents2022} can enhance the perceived safety of the system, thereby fostering trust among users. 
Reliability, on the other hand, refers to the consistent performance of the system. Interestingly, a system that is reliable and consistent, even if less accurate, can often be perceived as more trustworthy \citep{duanTransparentTrustworthyPrediction2023}. It may happen because users (i.e., teachers) can predict the system’s behaviour and adjust their expectations accordingly, increasing trust. This reliable system is important in empowering users by providing consistent and predictable results, enabling them to make informed decisions.
Moreover, trustworthiness is closely linked to both safety and reliability. A system that consistently protects user data (safety) and performs as expected (reliability) is likely to be deemed trustworthy. A trustworthy system empowers users by fostering confidence in the system’s outputs, which in turn encourages its adoption and use \citep{khosravi2022explainable}.

In conclusion, these principles are intertwined, and improvements in one area can positively impact the others \citep{ozmen2023six,shneidermanHCAI2020}. 
The interconnected principles of safety, reliability, and trustworthiness, coupled with a balance of human control and computer automation, can significantly empower stakeholders in the context of human-centred LA/AIED systems. 
Therefore, future research and development efforts should enhance these principles collectively to foster user trust and promote the acceptance and effectiveness of human-centred LA/AIED systems in real-world practices, accompanied by the aforementioned human-centred design practices above. Notably, beyond stakeholders' perspectives, we could not find studies exploring external mechanisms to ensure a human-centred LA/AIED system is safe, trustworthy or reliable. These mechanisms can involve, for example, subjecting the systems to independent audits and assessments by third parties to verify their actual trustworthiness, to move beyond perceived trust, or ensure adherence to relevant legal and regulatory requirements, as well as industry standards \citep{shneidermanHCAI2020}.

Therefore, we recommend:
% safety, reliability, and trustworthiness principles can enhance the credibility of human-centred LA/AIED systems.
\begin{itemize}
    \item \textbf{Safety Procedures}: Report safety procedures, including risk assessment, stakeholders’ perspectives, and informed consent procedures.
    \item \textbf{Data Privacy}: Ensure data privacy of stakeholders when involving them in the design process. Include secure data storage procedures and risk assessments for safety in the educational environment.
    \item \textbf{Reliability \& Collaboration}: Foster collaboration with end-users and employ system explainability. Educate end-users on system capabilities and limitations to improve privacy awareness and effective technology use.
    \item \textbf{Trustworthiness}: Report stakeholders’ trust perceptions, ensure transparency of data outcomes and algorithms, and measure trustworthiness using human control and computer automation dimensions.
    \item \textbf{Accountability}: Explore stakeholder accountability in the design process, focusing on how teachers or students can be held accountable for their inputs in human-centred LA/AIED systems.
\end{itemize}

\subsection{Limitations}
This SLR is subject to limitations that should be considered in interpreting results and discussions. First, despite the comprehensive search strategy employed, there is always a possibility of missing relevant studies due to limitations in the search terms, databases searched, or the exclusion of grey literature (e.g., unpublished works) and groundbreaking results in short articles or posters. Although we took many studies to reduce bias, it is possible that the study selection process could be influenced by the reviewers' judgment or subjective criteria, which could introduce some selection bias. However, this is a known possibility and has been addressed through transparency and open communication between researchers.
We further acknowledge that LA and AIED come from two different research communities and potentially different ways of reporting their research \citep{rientiesDefiningBoundariesArtificial2020}. Nonetheless, we found a clear overlap between these two communities. This SLR attempted to comprehensively view these overlaps through a human-centred perspective, allowing future work to explore and advance this avenue.

\section{Concluding remarks}

The findings of this systematic literature review indicate a growing interest in human-centred LA and AIED research. 
The review highlights the importance of safety, reliability, and trust in the design and implementation of these data-intensive systems, as well as the need for transparent, effective communication and user control. 
The review also identifies gaps in the existing research and methodological challenges that need to be addressed for human-centred LA and AIED to remain relevant and potentially become part of mainstream practices in the foreseeable future. 
Overall, this review provides valuable insights into the current state of human-centredness in LA and AIED studies with support HCAI framework as a lens, and underscores the importance of ongoing research and development.

%TC:ignore
\section*{Acknowledgements}
% % Funding: Hidden for double-blinded review.
Riordan Alfredo gratefully acknowledges Monash University for his PhD scholarship. This research was funded partially by the Australian Government through the Australian Research Council (project number DP210100060).

%% Loading bibliography style file
% \bibliographystyle{model1-num-names}
\bibliographystyle{cas-model2-names}

% Loading bibliography database
\bibliography{bib-intro, bib-results}

\onecolumn
\appendix
\section{HCAI two-dimensional quadrants in LA/AIED references}
\begin{longtable}{|c|p{1.5cm}|p{13cm}|}
    \hline
    \textbf{No.} & \textbf{Quadrant} & \textbf{Reference} \\
    \hline
\citenum{ahnDesigningContextReaching2019}& Q2 & Ahn, J., Campos, F., Hays, M., \& Digiacomo, D. (2019). Designing in context: Reaching beyond usability in learning analytics dashboard design. Journal of Learning Analytics, 6(2), 70–85. https://doi.org/10.18608/jla.2019.62.5                                                                      \\
 \citenum{al-doulatMakingSenseStudent2020}& Q2 & Al-Doulat, A., Nur, N., Karduni, A., Benedict, A., Al-Hossami, E., Maher, M. L., Dou, W., Dorodchi, M., \& Niu, X. (2020). Making Sense of Student Success and Risk Through Unsupervised Machine Learning and Interactive Storytelling. In Artificial Intelligence in Education (Vol. 12163, pp. 3–15). Springer International Publishing. http://link.springer.com/10.1007/978-3-030-52237-7\_1                                                   \\
 \citenum{alboKnowledgeBasedDesignAnalytics2022}& Q4 & Albó, L., Barria-Pineda, J., Brusilovsky, P., \& Hernández-Leo, D. (2022). Knowledge-Based Design Analytics for Authoring Courses with Smart Learning Content. Int. J. Artif. Intell. Educ., 32(1), 4–27. https://doi.org/10.1007/s40593-021-00253-3                                                                                                                                                                                               \\
 \citenum{alevenDashboardSupportTeachers2022}& Q4 & Aleven, V., Blankestijn, J., Lawrence, L., Nagashima, T., \& Taatgen, N. (2022). A Dashboard to Support Teachers During Students’ Self-paced AI-Supported Problem-Solving Practice. 16–30.                                                                                                                                                                                                                                                         \\
\end{longtable}
%TC:endignore 

\end{document}